\documentclass[12pt]{article}
\usepackage{amsmath}
\usepackage{amssymb}
\usepackage{amsfonts}
\usepackage{graphicx,psfrag,epsf}
\usepackage{natbib}
\usepackage[utf8]{inputenc}
 
\usepackage{hyperref}
\usepackage{mathtools}
\usepackage{threeparttable}
\usepackage{booktabs}
\usepackage{bookmark}
 
\usepackage{url} 
\usepackage{appendix}
 
\usepackage[figuresright]{rotating}

\usepackage{amsthm}
 
\def\bSig\mathbf{\Sigma}

\newtheorem{theorem}{Theorem}

\begin{document}


  \title{\bf Variance estimation in pseudo-expected estimating equations for missing data}
  \author{Giorgos Bakoyannis \\
    Department of Biostatistics and Health Data Science \\
     Indiana University Fairbanks School of Public Health \\ 
     \\
     Philani B. Mpofu \\
     Flatiron Health \\
     \\
     Andrea Broyles \\
     Center for Biomedical Informatics \\
     Regenstrief Instiatute Inc \\
     \\
     Brian E. Dixon \\
     Department of Epidemiology \\
     Indiana University Fairbanks School of Public Health \\
     Center for Biomedical Informatics \\
     Regenstrief Instiatute Inc} 
  \date{}
  \maketitle

\def\spacingset#1{\renewcommand{\baselinestretch}%
{#1}\small\normalsize} \spacingset{1}


\begin{abstract}
Missing data is a common challenge in biomedical research. This fact, along with growing dataset volumes of the modern era, make the issue of computationally-efficient analysis with missing data of crucial practical importance. A general computationally-efficient estimation framework for dealing with missing data is the pseudo-expected estimating equations (PEEE) approach. The method is applicable with any parametric model for which estimation involves the solution of a set of estimating equations, such as likelihood score equations. A key limitation of the PEEE approach is that there is currently no closed-form variance estimator, and variance estimation requires the computationally burdensome bootstrap method. In this work, we address the gap and provide a closed-form variance estimator whose computation can be significantly faster than a bootstrap approach. Our variance estimator is shown to be consistent even with auxiliary variables and under misspecified models for the incomplete variables. Simulation studies show that our variance estimator performs well and that its computation can be over 50 times faster than the bootstrap. The computational efficiency gain from our proposed variance estimator is crucial with large datasets or when the main analysis method is computationally intensive. Finally, the PEEE approach along with our variance estimator are used to analyze incomplete electronic health record data of patients with traumatic brain injury.
\end{abstract}

\noindent%
{\it Keywords:} Auxiliary variable; Imputation; Large dataset; Missing at random; Uncongeniality.
\vfill

\newpage
\spacingset{1.45} 

\section{Introduction}
\label{sec1}

The volume of data continues to grow at exponential rates. Estimates suggest that approximate 175 Zetabytes will be generated annually by 2025 \citep{Coughlin18}. 
However, large datasets are rarely complete and typically contain missing values in one or more variables. An example is electronic health records, which contain a significant frequency of incomplete information (4.3-86\%) about patients in real-world settings. Electronic health records are increasingly used in large, observational studies to produce real-world study of outcomes \citep{Dixon20} and develop predictive models for use in clinical applications \citep{Jones22}. Discarding observations with missing values can seriously threaten the validity of statistical inference and prediction based on such datasets, and can also lead to loss of statistical efficiency \citep{Little2002,Carpenter2012}. Therefore, dealing with missing data in a computationally-efficient manner is an important practical issue, especially in biomedical research.

In this paper, we consider parametric approaches for dealing with missingness due to their practical utility and computational efficiency, which are important in modern applications with large datasets. 
A popular approach for the analysis of incomplete datasets under a missing at random (MAR) assumption is Rubin's multiple imputation \citep{Rubin2004,Carpenter2012}. This imputation approach is also known as type A multiple imputation \citep{Wang98}. The main attractive features of Rubin's multiple imputation is that it can be easily implemented using off-the-shelf statistical software and is, in general, computationally efficient. 
An important limitation of type A multiple imputation is that Rubin's variance estimator is biased when the imputation model is misspecified \citep{Robins00,Wang98}, when the imputation and analysis models are uncongenial \citep{Meng94,Robins00,Wang98}, or when the analysis model is not self-efficient \citep{Meng94,Meng03,Yang16}. A common reason for the uncongeniality between the analysis and imputation models is the inclusion of auxiliary variables in the latter model, in an effort to make the MAR assumption more plausible in applications \citep{Lu01, Nevo18, Bakoyannis19}.

Since the introduction of Rubin's multiple imputation, there have been efforts to address the bias issue of Rubin's variance estimator. \citet{Kim11} developed an alternative multiple imputation approach called parametric fractional imputation. The proposed variance estimator for this approach relaxes the self-efficiency assumption in Rubin's multiple imputation but not the congeniality assumption \citep{Yang16}. However, the implementation of the parametric fractional imputation approach is more complicated and not as computationally efficient as Rubin's multiple imputation, requiring the use of an EM algorithm and the Newton method, which may limit its use in applications with large datasets. \citet{Wang98} developed the frequentist type B multiple imputation framework which enjoys the computational efficiency and simplicity of Rubin's type A multiple imputation, while providing consistent variance estimation even under a misspecified imputation model and uncongeniality between the analysis and imputation models. Moreover, type B multiple imputation is more statistically efficient than type A multiple imputation. Type B multiple imputation requires performing the complete data analysis on a stacked dataset that contains all the imputed datasets \citep{Robins00}. Letting $n$ be the total sample size and $S$ the number of imputations, the resulting stacked dataset is of size $nS$ \citep{Robins00}.

One characteristic of all the aforementioned multiple imputation approaches is that imputing the missing data finitely many times ($S$) leads to an additional source of variability. Nevertheless, this additional variability vanishes as $S\rightarrow \infty$. In practice, it is suggested to perform 10--20 imputations in order to limit the variability due to the finite number of imputations. However, analyzing a stacked dataset that is 10 to 20 times as large as the original dataset can be computationally burdensome with modern large datasets. Moreover, handling such large datasets in the R language, which utilizes the computer's RAM, can be challenging. An alternative general framework for missing data, which avoids multiple stochastic imputations, is the expected estimating equations (EEE) framework \citep{Wang08}. The EEE framework solves a set of expected, with respect to the missing variables, estimating equations. However, this approach estimates the parameters of interest and the nuisance parameters associated with the distribution of missing data simultaneously and, thus, it can be computationally intensive. For this reason, \citet{Wang08} considered in a couple of examples a computationally faster alternative which they called pseudo-expected estimating equations (PEEE) approach. Unfortunately, to the very best of our knowledge, no closed-form variance estimator has been proposed for this approach and variance estimation relies on bootstrap methods. The latter methods can be quite computationally burdensome with large datasets or when the main analysis method (i.e., the method for the analyzing complete data) is computationally intensive.

Our motivation comes from a study based on electronic health record data of 88,168 individuals with traumatic brain injury (TBI) who were hospitalized within 24 hours of their diagnosis. The scientific interest in this study is to evaluate predictors of rehabilitation within 3 months from the hospitalization date. A challenge in this dataset is that the race variable (coded as a three-level categorical variable) is missing for 25.5\% of the patients. This frequency of missing data is common for large electronic health record datasets, and is actually lower than that of some types of data often needed or used in biomedical research \citep{Balas15,Kharrazi14,Dixon13}. Moreover, it is expected that the probability of missingness is associated with the calendar year of admission. Given that one of the goals in this study is to use the fitted model to predict future cases of rehabilitation, it is not desirable to include calendar year as a covariate in the main analysis model. Therefore, calendar year is an auxiliary covariate in this case and, thus, type A multiple imputation is not the optimal analytical choice. Moreover, using type B multiple imputation in this example would lead to a somewhat computationally burdensome analysis as it would require to analyze a stacked dataset consisting of 881,680 observations, assuming 10 imputations. Using the PEEE approach, it can be shown using the arguments provided in Section \ref{PEEE_comp}, that this analysis can be performed by fitting a weighted logistic regression model to an augmented dataset of only 133,056 observations. Therefore, the PEEE approach can lead to a substantially faster computation of point estimates. Moreover, in this case, the PEEE estimator avoids additional variability due to the finite number of imputations in type B multiple imputation.

In this manuscript we address the issue of closed-form variance estimation in the PEEE approach for missing data problems under MAR. To achieve this, we rigorously establish, for the first time, the consistency and asymptotic normality of the PEEE estimator. The latter relies on modern empirical process theory \citep{van96, Kosorok08} which requires weaker assumptions than more traditional asymptotic theory techniques (more details on this are included in the Discussion section). From a practical standpoint, this means that our results are applicable to a broader range of settings. Our variance estimator is shown to be consistent even under a misspecified model for the missing data and under uncongeniality between this model and the main analysis model, unlike Rubin's variance formula. In addition, we present  details on how to compute the PEEE estimator using standard software that allows for case weights (see Appendix B). Simulation studies show that our variance estimator performs well even under uncongeniality and misspecification, and that its computation can be significantly faster than the bootstrap method for variance estimation. The latter is very important in modern applications with large datasets and/or computationally intensive main analysis methods. The PEEE approach and the proposed variance estimator are used to analyze our motivating TBI data. 


\section{Methods}
\label{sec2}
\subsection{Notation and Data}
Let $X\in\mathbb{R}^p$ be a vector that includes both dependent and independent variables. When some variables are incompletely observed, this vector can be partitioned as $X=(X^o,X^m)$, where $X^{o}$ denotes the fully observed portion of $X$ and $X^{m}$ the incompletely unobserved subset of $X$. $X^m$ may include a dependent variable and/or covariates, depending on the setting. For simplicity of presentation and without loss of generality we consider the case where only one variable is incompletely observed. Let $R$ denote the missingness indicator, where $R=1$ if $X^m$ has been observed and $R=0$ if the value of $X^m$ is missing. Along with $X$, one typically also observes a vector of fully observed auxiliary variables $A=(A_{1},\ldots,A_{q})'$, which is not of direct scientific interest but may be related to the probability of missingness. It has to be noted that incorporation of auxiliary variables can make the MAR assumption more plausible in applications \citep{Lu01,Nevo18,Bakoyannis19}. The data based on $n$ i.i.d. observations are $D_i=(R_i,A_i,X_i^o,R_iX_i^m)$, $i=1,\ldots,n$. 

\subsection{PEEE Approach}
Suppose that the scientific interest lies in estimating the parameter $\theta_0\in\Theta\subset\mathbb{R}^d$ in a parametric model, based on the sample of $n$ i.i.d. observations. Typically, if there are no missing data, estimation of $\theta_0$ can be performed by solving an  estimating equation of the form
\[
\sum_{i=1}^n\psi_{\hat{\theta}_n}(X_i)=0.
\]
In parametric maximum likelihood problems, $\psi_{\theta}(X_i)$ is the vector of partial derivatives of the loglikelihood contribution of the $i$th observation. Note that estimation of $\theta_0$ can be equivalently performed by solving
\begin{equation}
\Psi_n(\hat{\theta}_n)\equiv \frac{1}{n}\sum_{i=1}^n\psi_{\hat{\theta}_n}(X_i)=0, \label{ee}
\end{equation}
since multiplication by $1/n$ does not affect the solution of this equation.

With missing data, $\psi_{\theta}(X_i)=\psi_{\theta}(X_i^o,X_i^m)$ cannot be evaluated for the incomplete observations.  For such cases, it possible to replace $\psi_{\theta}(X_i^o,X_i^m)$ by the conditional expectation $E[\psi_{\theta}(X^o,X^m)|X_i^o,A_i]$, which only depends on the fully observed variables $X^o$ and $A$. The rationale for this is grounded on the fact that, with estimating equations, $\Psi_n(\theta)$ is an estimate of
\[
\Psi(\theta)\equiv E[\psi_{\theta}(X)]
\]
for which $\Psi(\theta_0)=0$ for the true parameter $\theta_0$ only (by the identifiability of the model), along the fact that
\[
E[\psi_{\theta}(X)]=E\{E[\psi_{\theta}(X)|X^o,A]\}.
\]
Therefore, a reasonable estimate of $\Psi(\theta)$ is
\[
\frac{1}{n}\sum_{i=1}^n \left\{R_i\psi_{\theta}(X_i^o,X_i^m)+(1-R_i)E[\psi_{\theta}(X^o,X^m)|X^o,A]\right\},
\]
if the data are MAR. The MAR assumption imposed here is:
\begin{equation}
P(R=1|X^m,X^o,A)=P(R=1|X^o,A). \label{mar}
\end{equation}
The MAR assumption in \eqref{mar} is weaker than the usual MAR assumption of independence between $R$ and $X^m$ conditionally on the observed data $X^o$ only. Assumption \eqref{mar} is equivalent to
\begin{eqnarray}
P(X^m|R=1,X^o,A)&=&P(X^m|R=0,X^o,A) \nonumber \\
&=& P(X^m|X^o,A). \label{mar2}
\end{eqnarray}
In practice, a parametric model $F_{X^m|X^o,A}(x|x^o,a;\gamma_0)$ is typically assumed for the incomplete variable $X^m$, which depends on an unknown parameter $\gamma_0$. It is not hard to see that, under the latter assumption, $E[\psi_{\theta}(X^o,X^m)|X^o,A]$ also depends on $\gamma_0$. Therefore, the expected estimating equations can be expressed as
\begin{eqnarray}
\tilde{\Psi}_n(\theta;\gamma_0)&\equiv&
\frac{1}{n}\sum_{i=1}^n \left\{R_i\psi_{\theta}(X_i^o,X_i^m)+(1-R_i)E[\psi_{\theta}(X^o,X^m)|X_i^o,A_i;\gamma_0]\right\} \nonumber \\
&\equiv&\frac{1}{n}\sum_{i=1}^n\tilde{\psi}_{\theta,\gamma_0}(D_i). \label{eee}
\end{eqnarray}
Estimation of $\theta_0$ can be performed in a two-stage fashion. The first stage of the analysis involves the computation of an estimate $\hat{\gamma}_n$ of $\gamma_0$ in \eqref{eee} using the complete observations in light of the MAR assumption \eqref{mar2}. The second stage of the analysis proceeds with replacing the unknown $\gamma_0$ by $\hat{\gamma}_n$ in the estimating function and solving the estimating equation $\tilde{\Psi}_n(\hat{\theta}_n;\hat{\gamma}_n)=0$ to estimate $\theta_0$. The resulting estimator is the PEEE estimator as it involves solving a set of expected estimating equations that depends on the estimated parameter $\hat{\gamma}_n$. 

\subsection{A Note on Computation}
\label{PEEE_comp}
The PEEE estimator can be easily obtained using any standard software that allows for case weights using a simple data augmentation technique. This computation technique is illustrated in detail in Appendix B. If the incomplete variable is linear in $\psi_{\theta}(X^o,X^m)$, then the conditional expectation in $\tilde{\Psi}_n(\theta;\hat{\gamma}_n)$ can be computed as 
\[
E[\psi_{\theta}(X^o,X^m)|X_i^o,A_i;\hat{\gamma}_n]=\psi_{\theta}\left(X_i^o,E(X^m|X_i^o,A_i;\hat{\gamma}_n)\right),
\]
and the problem reduces to a single deterministic imputation of the missing $X_i^m$ with $E(X^m|X_i^o,A_i;\hat{\gamma}_n)$, under the assumed $F_{X^m|X^o,A}(x|x^o,a;\gamma_0)$. If the incomplete variable is discrete and takes its values in the finite set $\{x_1,\ldots,x_K\}$, then the conditional expectation can be computed as 
\[
E[\psi_{\theta}(X^o,X^m)|X_i^o,A_i;\hat{\gamma}_n]=\sum_{k=1}^K\psi_{\theta}\left(X_i^o,x_k\right)P(X^m=x_k|X_i^o,A_i;\hat{\gamma}_n).
\]
When the conditional expectation can be calculated analytically, the PEEE approach avoids the additional variability due to the finite number of stochastic imputations in both Type A and Type B multiple imputation. If the conditional expectation cannot be calculated analytically, then one can utilize Monte Carlo methods to calculate the conditional expectation in $\tilde{\Psi}_n(\theta;\hat{\gamma}_n)$. Letting $S$ be the number of simulation realizations of the incomplete variables from $F_{X^m|X^o,A}(x|X_i^o,A_i;\hat{\gamma}_n)$ and $U_{ij}$, $j=1,\ldots,S$ be independent realizations from the uniform $U(0,1)$ distribution, the Monte Carlo version of the pseudo-expected estimating function is
\begin{eqnarray*}
\tilde{\Psi}_{n,S}(\theta;\hat{\gamma}_n)&\equiv&
\frac{1}{n}\sum_{i=1}^n \left\{R_i\psi_{\theta}(X_i^o,X_i^m)+(1-R_i)\frac{1}{S}\sum_{j=1}^S\psi_{\theta}\left(X_i^o,\tilde{X}_i^m(U_{ij},\hat{\gamma}_n)\right)\right\} \\
&\equiv&\frac{1}{n}\sum_{i=1}^n\tilde{\psi}_{\theta,\hat{\gamma}_n,S}(D_i), 
\end{eqnarray*}
where $\tilde{X}_i^m(U_{ij},\hat{\gamma}_n)=F_{X^m|X^o,A}^{-1}(U_{ij}|X_i^o,A_i;\hat{\gamma}_n)$, provided that the latter inverse exists. We call the estimator $\hat{\theta}_{n,S}$ that satisfies $\tilde{\Psi}_{n,S}(\hat{\theta}_{n,S};\hat{\gamma}_n)=0$ the Monte Carlo PEEE (MCPEEE) estimator. Given that $\tilde{X}_i^m(U_{ij},\hat{\gamma}_n)$ is a simulation realization from the distribution $F_{X^m|X^o,A}(x|X_i^o,A_i;\hat{\gamma}_n)$, the Monte Carlo PEEE approach is equivalent to type B multiple imputation. However, its computation requires  utilizing a stacked dataset of only $n+m(S-1)$ observations, where $m$ is the number of incomplete observations, instead of a stacked dataset of size $nS$ for the type B multiple imputation approach. This is expected to lead to substantial computational efficiency gains.

\section{Asymptotic Properties and Variance Estimators}
\label{sec3}
Assuming that the MAR assumption holds, the parametric model imposed for the incomplete variable is correctly specified, and some other weak regularity conditions listed in Appendix A are satisfied, we establish the asymptotic properties of the PEEE estimator $\hat{\theta}_n$ and the MCPEEE estimator $\hat{\theta}_{n,S}$. We also propose closed-form variance estimators for both $\hat{\theta}_n$ and $\hat{\theta}_{n,S}$.  
\begin{theorem}
Suppose conditions C1--C7$^{\prime}$ listed in Appendix A hold. Then, 
\[
\|\hat{\theta}_n-\theta_0\|\overset{p}\rightarrow 0 \ \ \textrm{and} \ \ \|\hat{\theta}_{n,S}-\theta_0\|\overset{p}\rightarrow 0,
\]
as $n\rightarrow\infty$.
\end{theorem}
The proof of Theorem 1 is outlined in Appendix A.1. Note that the MCPEEE estimator is consistent for any choice of $S$ (even with $S=1$). This phenomenon is similar to the consistency of the equivalent type B multiple imputation estimator for any choice of $S\in\{1,2,\ldots\}$. Below, we will argue that a larger $S$ gives a more precise estimate at the cost of additional computation time. Before providing the result for the asymptotic distribution of $\hat{\theta}_n$ we define the fixed quantities
\[
\dot{\Psi}_{\theta_0}=\frac{\partial}{\partial\theta}\Psi(\theta)\big|_{\theta=\theta_0}=\frac{\partial}{\partial\theta}E[\tilde{\psi}_{\theta,\gamma_0}(D)]\big|_{\theta=\theta_0},
\]
and
\[
G_{\theta_0,\gamma_0}=E\left\{(1-R)\frac{\partial}{\partial\gamma}E[\psi_{\theta_0}(X^o,X^m)|X^o,A;\gamma]\big|_{\gamma=\gamma_0}\right\}.
\]
Also, let $\omega_i$ be the influence function of the $i$th observation for the estimator $\hat{\gamma}_n$. If $\hat{\gamma}_n$ is a maximum likelihood estimate then $\omega_i=I^{-1}(\gamma_0)U_i(\gamma_0)$, where $I(\gamma_0)$ is the information matrix for $\gamma_0$ for a single observation and $U_i(\gamma_0)$ the score function contribution of the $i$th observation. Next, define the empirical versions of $\dot{\Psi}_{\theta_0}$ and $G_{\theta_0,\gamma_0}$ as
\[
\dot{\Psi}_{n,\hat{\theta}_n}(\hat{\gamma}_n)=\frac{1}{n}\sum_{i=1}^n\frac{\partial}{\partial\theta}\tilde{\psi}_{\theta,\hat{\gamma}_n}(D_i)\big|_{\theta=\hat{\theta}_n},
\]
and
\[
\hat{G}_{\hat{\theta}_n,\hat{\gamma}_n}=\frac{1}{n}\sum_{i=1}^n(1-R_i)\frac{\partial}{\partial\gamma}E[\psi_{\hat{\theta}_n}(X^o,X^m)|X_i^o,A_i;\gamma]\big|_{\gamma=\hat{\gamma}_n}.
\]
The empirical versions of the influence functions $\omega_i$, $i=1,\ldots,n$, can be obtained by replacing the unknown parameters with their corresponding consistent estimates and the expectations by sample averages. For the case where $\hat{\gamma}_n$ is a maximum likelihood estimator $\hat{\omega}_i=\hat{I}_n^{-1}(\hat{\gamma}_n)U_i(\hat{\gamma}_n)$, where $\hat{I}_n^{-1}(\hat{\gamma}_n)$ is the empirical information matrix evaluated at the estimated parameter $\hat{\gamma}_n$.

\begin{theorem}
Suppose conditions C1--C7 listed in Appendix A hold. Then
\[
\sqrt{n}(\hat{\theta}_n-\theta_0)\overset{d}\rightarrow N(0,\Omega),
\]
as $n\rightarrow\infty$, where
\[
\Omega=\left(\dot{\Psi}_{\theta_0}^{-1}\right)E\left[\tilde{\psi}_{\theta_0,\gamma_0}(D)+G_{\theta_0,\gamma_0}\omega\right]^{\otimes 2}\left(\dot{\Psi}_{\theta_0}^{-1}\right),
\]
with $B^{\otimes 2}=BB^T$ for a matrix $B$. Moreover,
\[
\hat{\Omega}_n=\dot{\Psi}_{n,\hat{\theta}_n}^{-1}(\hat{\gamma}_n)\left\{\frac{1}{n}\sum_{i=1}^n\left[\tilde{\psi}_{\hat{\theta}_n,\hat{\gamma}_n}(D_i)+\hat{G}_{\hat{\theta}_n,\hat{\gamma}_n}\hat{\omega}_i\right]^{\otimes 2}\right\}\dot{\Psi}_{n,\hat{\theta}_n}^{-1}(\hat{\gamma}_n)\overset{p}\rightarrow\Omega.
\]
\end{theorem}
The proof of Theorem 2 is provided in Appendix A.2. It is important to note that, unlike Rubin's variance formula for type A multiple imputation, the proposed variance estimator $\hat{\Omega}_n$ is consistent even under uncongeniality (e.g., due to the presence of auxiliary variables $A$ in the model for the incomplete variable $X^m$) and a misspecified model $F_{X^m|X^o,A}(x|x^o,a;\gamma_0)$. When the complete-data analysis relies on maximum likelihood estimation, $\dot{\Psi}_{n,\theta_n}^{-1}(\hat{\gamma}_n)$ is equal to the na\"ive (non-robust) variance matrix (i.e. the inverse of the estimated information matrix) of the main analysis model provided by default by standard packages, multiplied by the sample size $n$. Also, $\tilde{\psi}_{\hat{\theta}_n,\hat{\gamma}_n}(D_i)$ are the individual score contributions for the main analysis model, which are typically provided by standard packages. In R, for example, one can use the function \texttt{estfun} of the package \texttt{sandwich} to get these score contributions. Similarly, if $\hat{\gamma}_n$ is a maximum likelihood estimator, one can calculate $\hat{\omega}_i=\hat{I}_n^{-1}(\hat{\gamma}_n)U_i(\hat{\gamma}_n)$. Finally, calculation of $\hat{G}_{\hat{\theta}_n,\hat{\gamma}_n}$ requires the estimates $\hat{\theta}_n$ and $\hat{\gamma}_n$ along with the derivative of $E[\psi_{\hat{\theta}_n}(X^o,X^m)|X^o,A;\gamma]$ with respect to $\gamma$ evaluated at $\hat{\gamma}_n$. This derivative can be calculated either analytically or numerically. 

A key assumption for the consistency of $\hat{\theta}_n$, as for any other parametric imputation-type estimator for missing data, is that the parametric model $F_{X^m|X^o,A}(x|x^o,a;\gamma_0)$ is correctly specified (condition C5 in Appendix A). This assumption can be relaxed a bit in the PEEE estimator by assuming a correct specification of just the conditional mean model $E[\psi_{\theta}(X^o,X^m)|X^o=x^o,A=a;\gamma_0]$ and not the full conditional distribution $F_{X^m|X^o,A}(x|x^o,a;\gamma_0)$, as long as the parameter $\gamma_0$ can be consistently estimated (condition C6 in Appendix A). 
In contrast, both type A and type B multiple imputation estimators require full specification of the conditional distribution of the incomplete variable given the observed variables in order to simulate the missing values. We must note that the same is true for the MCPEEE estimator $\hat{\theta}_{n,S}$ that is used in cases where the conditional expectation cannot be specified in closed form.

When the assumed model $F_{X^m|X^o,A}(x|x^o,a;\gamma_0)$ (or $E[\psi_{\theta}(X^o,X^m)|X^o=x^o,A=a;\gamma_0]$) is misspecified, it can be shown that $\hat{\gamma}_n\rightarrow\gamma^*\neq\gamma_0$ and (under similar conditions and using similar arguments to those used in Appendix A) that the PEEE estimator converges to some value $\theta^*\neq\theta_0$. When $\hat{\gamma}_n$ is a maximum likelihood estimator, $\gamma^*$ is the minimizer of the Kullback--Leibler divergence between the assumed model and the true distribution $F_{X^m|X^o,A}(x|x^o,a)$ (or the true conditional mean $E[\psi_{\theta}(X^o,X^m)|X^o=x^o,A=a]$). In such cases, $\theta^*$ is the solution to the (true) expected estimating equations under the \textit{closest} (with respect to the Kullback--Leibler divergence) model to the true distribution $F_{X^m|X^o,A}(x|x^o,a)$ (or the true conditional mean $E[\psi_{\theta}(X^o,X^m)|X^o=x^o,A=a]$). To alleviate the consequences of misspecifying the model for $X^m$, one can use a flexible parametric model including, for example, regression splines. Regardless of whether $\hat{\gamma}_n$ is a maximum likelihood estimate or not, it can be shown using similar conditions and arguments to those used in the proof of Theorem 2 that, under misspecification,
\[
\sqrt{n}(\hat{\theta}_n-\theta^*)\overset{d}\rightarrow N(0,\Omega^*),
\]
as $n\rightarrow\infty$, where
\[
\Omega^*=\left(\dot{\Psi}_{\theta^*}^{-1}(\gamma^*)\right)E\left[\tilde{\psi}_{\theta^*,\gamma^*}(D)+G_{\theta^*,\gamma^*}\omega^*\right]^{\otimes 2}\left(\dot{\Psi}_{\theta^*}^{-1}(\gamma^*)\right),
\]
with
\[
\dot{\Psi}_{\theta^*}(\gamma^*)=\frac{\partial}{\partial\theta}E[\tilde{\psi}_{\theta,\gamma^*}(D)]\big|_{\theta=\theta^*}
\]
and $\omega_i^*=I^{-1}(\gamma^*)U_i(\gamma^*)$. Moreover, the estimator provided in Theorem 2 is consistent, that is $\hat{\Omega}_n\overset{p}\rightarrow\Omega^*$. Therefore, similarly to the case of type B multiple imputation \citep{Wang98,Robins00}, the proposed variance estimator is consistent even under misspecification of the model for the incomplete variable $X^m$ and under uncongeniality between this model and the main analysis model.

For the asymptotic distribution of the MCPEEE estimator we need to introduce the following fixed quantity:
\[
H_{\theta_0,\gamma_0}=E\left[(1-R)\frac{\partial}{\partial\gamma}\psi_{\theta_0}(X^o,F_{X^m|X^o,A}^{-1}(U|X^o,A;\gamma))\big|_{\gamma=\gamma_0}\right].
\]
The empirical version of the latter quantity is
\[
\hat{H}_{\hat{\theta}_n,\hat{\gamma}_n}=\frac{1}{n}\sum_{i=1}^n\left[(1-R_i)\frac{1}{S}\sum_{j=1}^S\frac{\partial}{\partial\gamma}\psi_{\hat{\theta}_0}(X_i^o,F_{X^m|X^o,A}^{-1}(U_{ij}|X_i^o,A_i;\gamma))\big|_{\gamma=\hat{\gamma}_n}\right].
\]
\begin{theorem}
Suppose conditions C1--C6 and C7$^{\prime}$ listed in Appendix A hold. Then
\[
\sqrt{n}(\hat{\theta}_{n,S}-\theta_0)\overset{d}\rightarrow N(0,\Omega_S),
\]
as $n\rightarrow\infty$, where
\[
\Omega_S=\left(\dot{\Psi}_{\theta_0}^{-1}\right)P\left(\tilde{\psi}_{\theta_0,\gamma_0,S}+H_{\theta_0,\gamma_0}\omega\right)^{\otimes 2}\left(\dot{\Psi}_{\theta_0}^{-1}\right).
\]
Moreover,
\[
\hat{\Omega}_{n,S}=\dot{\Psi}_{n,\hat{\theta}_n}^{-1}(\hat{\gamma}_n)\left\{\frac{1}{n}\sum_{i=1}^n\left[\tilde{\psi}_{\hat{\theta}_n,\hat{\gamma}_n,S}(D_i)+\hat{H}_{\hat{\theta}_n,\hat{\gamma}_n}\hat{\omega}_i\right]^{\otimes 2}\right\}\dot{\Psi}_{n,\hat{\theta}_n}^{-1}(\hat{\gamma}_n)\overset{p}\rightarrow\Omega_S.
\]
\end{theorem}
The proof of Theorem 3 is provided in Appendix A.3. Note that the MCPEEE estimator $\hat{\theta}_{n,S}$ typically exhibits a larger variability compared to the PEEE estimator $\hat{\theta}_{n}$. This is attributed to the additional variability of the standard uniform variables in $\tilde{\psi}_{\theta_0,\gamma_0,S}(D)$ due to the finite number $S$ of imputations. Increasing the number $S$ leads to a more precise estimate $\hat{\theta}_{n,S}$ at the cost of additional computation time. In applications with large datasets, where computational efficiency is more crucial than statistical efficiency, it may be reasonable to set $S=5$. Similarly to our variance estimator for the PEEE estimator, $\hat{\Omega}_{n,S}$ is a consistent estimator of the asymptotic variance of $\hat{\theta}_{n,S}$ even under a misspecified model $F_{X^m|X^o,A}(x|x^o,a;\gamma_0)$ for the incomplete variable and under uncongeniality between this model and the main analysis model.

\section{Simulation Studies}
\label{sec4}
We conducted a series of simulation experiments to evaluate numerically the performance of our proposed variance estimator for the PEEE approach with missing data. We considered two covariates of interest; $Z_1$ was a continuous covariate simulated from the standard normal distribution while $Z_2$ was a categorical covariate with three levels, which was simulated based on the probabilities $P(Z_2=1)=0.5$, $P(Z_2=2)=0.3$ and $P(Z_2=3)=0.2$. Similarly to our motivating study, we considered a binary dependent variable, denoted as $Y$, in the simulation experiments. For this variable we assumed the logistic model
\[
\textrm{logit}[P(Y=1|Z)]=\beta_0 + \beta_1Z_1 +\beta_2I(Z_2=2) + \beta_3I(Z_2=3),
\]
where $(\beta_0,\beta_1,\beta_2,\beta_3)=(-0.2,0.5,-0.75,0.25)$. We considered missingness in the three-level categorical covariate $Z_2$, to mimic our motivating study. The missingness indicator was simulated based on the logit model
\[
\textrm{logit}(P(R=1|A))=\eta + A,
\]
where $A$ was a continuous auxiliary variable generated as
\[
A=\log(1.5)+I(Z_2=2)-I(Z_2=3)-Y+\epsilon,
\]
with $\epsilon\sim N(0,1)$. The parameter $\eta$ was set equal to $-1.1$ or $-0.2$ depending on the scenario. These choices for $\eta$ led on average to 32.2\% and 48.1\% missingness, respectively. Under this simulation setup, the auxiliary variable $A$ needs to be taken into account in order to satisfy the MAR assumption. This is because $A$ is associated with both the probability of missingess and the incompletely observed variable $Z_2$. Under the simulation setup, the true model for $Z_2$ was
\[
\log\left[\frac{P(Z_2=j|Z_1,Y,A)}{P(Z_2=1|Z_1,Y,A)}\right]=\gamma_{j,0}+\gamma_{j,1}h_j(Z_1)+\gamma_{j,2}I(Y=1)+\gamma_{j,3}A, \ \ \ \ j=2,3,
\]
where
\[
\gamma_{j,0}=\log\left[\frac{P(Z_2=j)}{0.5}\right]-0.2-[I(j=3)-I(j=2)](0.5+\eta),
\]
$\gamma_{j,1}=-1$,
\[
h_j(Z_1)=\log\left\{\frac{1+\exp[-0.2+0.5Z_1- 0.75I(j=2) + 0.25I(j=3)]}{1+\exp(-0.2+0.5Z_1)}\right\},
\]
and $\gamma_{j,2}=\gamma_{j,3}=[I(j=3)-I(j=2)]$. Note that the true model was a multinomial logistic model which was nonlinear in $Z_1$. For both missingness scenarios we simulated 1000 datasets and considered the sample sizes $n\in\{1000,5000,10000\}$. In this simulation we evaluated the type B multiple imputation \citep{Wang98,Robins00} based on 5 and 20 imputations, and the PEEE approach. For both methods we assumed the multinomial logit model
\[
\log\left[\frac{P(Z_2=j|Z_1,Y,A)}{P(Z_2=1|Z_1,Y,A)}\right]=\gamma_{j,0}^*+\gamma_{j,1}^*Z_1+\gamma_{j,2}^*I(Y=1)+\gamma_{j,3}^*A, \ \ \ \ j=2,3.
\]
This model was misspecified since, unlike the true model listed above, it assumed a linear effect of $Z_1$. Standard error estimation was based on the nonparametric bootstrap with 100 replications for the type B multiple imputation approach, and on the proposed closed-form variance estimator for the PEEE approach. Results from the simulation studies are listed in Tables~\ref{sim1k}--\ref{sim10kb}.

\begin{table}
\caption{Simulation results for $n=1000$ and under 32.2\% missingness.}
\label{sim1k}
\centering
\begin{tabular}{ccccccc}
\toprule
Method & $\beta$ & Bias (\%) & MCSD & ASE & CP & RE \\
\hline
MIB(5)&$\beta_0$&2.256&0.104&0.104&0.949&1.023\\
&$\beta_1$&0.183&0.072&0.073&0.958&1.003\\
&$\beta_2$&0.333&0.201&0.202&0.955&1.046\\
&$\beta_3$&0.785&0.190&0.194&0.965&1.030\\[1ex]
MIB(20)&$\beta_0$&2.482&0.103&0.103&0.945&1.008\\
&$\beta_1$&0.144&0.072&0.073&0.956&1.001\\
&$\beta_2$&0.138&0.198&0.200&0.947&1.017\\
&$\beta_3$&0.782&0.188&0.193&0.960&1.009\\[1ex]
PEEE&$\beta_0$&2.605&0.103&0.103&0.949&1.000\\
&$\beta_1$&0.100&0.072&0.072&0.961&1.000\\
&$\beta_2$&0.065&0.197&0.196&0.953&1.000\\
&$\beta_3$&0.864&0.187&0.191&0.962&1.000\\
\bottomrule
\end{tabular}
\\
{MIB($S$): Type B multiple imputation based on $S$ imputations; MCSD: Monte Carlo standard deviation of the estimates; ASE: Average of the standard error estimates; CP: Empirical coverage probability of the 95\% confidence intervals; RE: Relative efficiency (larger values indicate larger variance compared to the variance of the PEEE approach)}
\end{table}

\begin{table}
\caption{Simulation results for $n=1000$ and under 48.1\% missingness.}
\label{sim1kb}
\centering
\begin{tabular}{ccccccc}
\toprule
Method & $\beta$ & Bias (\%) & MCSD & ASE & CP & RE \\
\hline
MIB(5)&$\beta_0$&0.388&0.118&0.115&0.940&1.050\\
&$\beta_1$&0.439&0.074&0.074&0.952&1.005\\
&$\beta_2$&1.683&0.245&0.239&0.942&1.054\\
&$\beta_3$&0.128&0.208&0.215&0.964&1.067\\[1ex]
MIB(20)&$\beta_0$&0.779&0.115&0.114&0.947&1.009\\
&$\beta_1$&0.413&0.074&0.074&0.958&1.001\\
&$\beta_2$&1.331&0.241&0.235&0.943&1.017\\
&$\beta_3$&0.292&0.202&0.214&0.966&1.006\\[1ex]
PEEE&$\beta_0$&0.933&0.115&0.113&0.945&1.000\\
&$\beta_1$&0.325&0.073&0.073&0.958&1.000\\
&$\beta_2$&1.204&0.239&0.230&0.939&1.000\\
&$\beta_3$&0.378&0.202&0.210&0.967&1.000\\
\bottomrule
\end{tabular}
\\
{MIB($S$): Type B multiple imputation based on $S$ imputations; MCSD: Monte Carlo standard deviation of the estimates; ASE: Average of the standard error estimates; CP: Empirical coverage probability of the 95\% confidence intervals; RE: Relative efficiency (larger values indicate larger variance compared to the variance of the PEEE approach)}
\end{table}

\begin{table}
\caption{Simulation results for $n=5000$ and under 32.2\% missingness.}
\label{sim5k}
\centering
\begin{tabular}{ccccccc}
\toprule
Method & $\beta$ & Bias (\%) & MCSD & ASE & CP & RE \\
\hline
MIB(5)&$\beta_0$&0.104&0.045&0.046&0.952&1.021\\
&$\beta_1$&-0.042&0.032&0.032&0.944&1.005\\
&$\beta_2$&0.089&0.085&0.089&0.955&1.049\\
&$\beta_3$&0.630&0.090&0.086&0.936&1.031\\[1ex]
MIB(20)&$\beta_0$&0.200&0.045&0.046&0.955&1.004\\
&$\beta_1$&-0.049&0.032&0.032&0.940&1.000\\
&$\beta_2$&-0.011&0.083&0.087&0.962&1.010\\
&$\beta_3$&0.877&0.088&0.085&0.935&1.004\\[1ex]
PEEE&$\beta_0$&0.261&0.045&0.046&0.958&1.000\\
&$\beta_1$&-0.059&0.032&0.032&0.946&1.000\\
&$\beta_2$&-0.043&0.083&0.087&0.960&1.000\\
&$\beta_3$&0.891&0.088&0.085&0.937&1.000\\
\bottomrule
\end{tabular}
\\
{MIB($S$): Type B multiple imputation based on $S$ imputations; MCSD: Monte Carlo standard deviation of the estimates; ASE: Average of the standard error estimates; CP: Empirical coverage probability of the 95\% confidence intervals; RE: Relative efficiency (larger values indicate larger variance compared to the variance of the PEEE approach)}
\end{table}

\begin{table}
\caption{Simulation results for $n=5000$ and under 48.1\% missingness.}
\label{sim5kb}
\centering
\begin{tabular}{ccccccc}
\toprule
Method & $\beta$ & Bias (\%) & MCSD & ASE & CP & RE \\
\hline
MIB(5)&$\beta_0$&0.599&0.049&0.051&0.960&1.037\\
&$\beta_1$&-0.027&0.032&0.032&0.950&1.002\\
&$\beta_2$&-0.330&0.104&0.105&0.954&1.046\\
&$\beta_3$&0.529&0.097&0.095&0.949&1.048\\[1ex]
MIB(20)&$\beta_0$&0.599&0.048&0.050&0.965&1.006\\
&$\beta_1$&-0.011&0.032&0.032&0.944&0.998\\
&$\beta_2$&-0.357&0.102&0.102&0.950&1.015\\
&$\beta_3$&0.659&0.095&0.094&0.951&0.997\\[1ex]
PEEE&$\beta_0$&0.663&0.048&0.050&0.964&1.000\\
&$\beta_1$&-0.030&0.032&0.032&0.945&1.000\\
&$\beta_2$&-0.399&0.101&0.102&0.950&1.000\\
&$\beta_3$&0.713&0.095&0.093&0.948&1.000\\
\bottomrule
\end{tabular}
\\
{MIB($S$): Type B multiple imputation based on $S$ imputations; MCSD: Monte Carlo standard deviation of the estimates; ASE: Average of the standard error estimates; CP: Empirical coverage probability of the 95\% confidence intervals; RE: Relative efficiency (larger values indicate larger variance compared to the variance of the PEEE approach)}
\end{table}

\begin{table}
\caption{Simulation results for $n=10000$ and under 32.2\% missingness.}
\label{sim10k}
\centering
\begin{tabular}{ccccccc}
\toprule
Method & $\beta$ & Bias (\%) & MCSD & ASE & CP & RE \\
\hline
MIB(5)&$\beta_0$&0.560&0.032&0.033&0.951&1.025\\
&$\beta_1$&0.035&0.022&0.023&0.948&1.009\\
&$\beta_2$&0.135&0.063&0.063&0.952&1.026\\
&$\beta_3$&1.232&0.061&0.061&0.951&1.021\\[1ex]
MIB(20)&$\beta_0$&0.562&0.032&0.032&0.950&1.011\\
&$\beta_1$&0.034&0.022&0.023&0.948&1.000\\
&$\beta_2$&0.209&0.063&0.062&0.949&1.015\\
&$\beta_3$&1.345&0.061&0.060&0.948&1.010\\[1ex]
PEEE&$\beta_0$&0.585&0.032&0.032&0.956&1.000\\
&$\beta_1$&0.031&0.022&0.023&0.951&1.000\\
&$\beta_2$&0.181&0.062&0.062&0.955&1.000\\
&$\beta_3$&1.353&0.061&0.060&0.956&1.000\\
\bottomrule
\end{tabular}
\\
{MIB($S$): Type B multiple imputation based on $S$ imputations; MCSD: Monte Carlo standard deviation of the estimates; ASE: Average of the standard error estimates; CP: Empirical coverage probability of the 95\% confidence intervals; RE: Relative efficiency (larger values indicate larger variance compared to the variance of the PEEE approach)}
\end{table}

\begin{table}
\caption{Simulation results for $n=10000$ and under 48.1\% missingness.}
\label{sim10kb}
\centering
\begin{tabular}{ccccccc}
\toprule
Method & $\beta$ & Bias (\%) & MCSD & ASE & CP & RE \\
\hline
MIB(5)&$\beta_0$&0.938&0.035&0.036&0.950&1.047\\
&$\beta_1$&0.043&0.023&0.023&0.944&1.011\\
&$\beta_2$&-0.127&0.073&0.074&0.958&1.048\\
&$\beta_3$&1.539&0.067&0.067&0.947&1.036\\[1ex]
MIB(20)&$\beta_0$&0.979&0.035&0.036&0.951&1.014\\
&$\beta_1$&0.049&0.023&0.023&0.945&1.000\\
&$\beta_2$&-0.069&0.072&0.072&0.954&1.010\\
&$\beta_3$&1.763&0.067&0.066&0.952&1.010\\[1ex]
PEEE&$\beta_0$&0.975&0.035&0.036&0.954&1.000\\
&$\beta_1$&0.043&0.023&0.023&0.952&1.000\\
&$\beta_2$&-0.093&0.071&0.072&0.961&1.000\\
&$\beta_3$&1.685&0.066&0.066&0.952&1.000\\
\bottomrule
\end{tabular}
\\
{MIB($S$): Type B multiple imputation based on $S$ imputations; MCSD: Monte Carlo standard deviation of the estimates; ASE: Average of the standard error estimates; CP: Empirical coverage probability of the 95\% confidence intervals; RE: Relative efficiency (larger values indicate larger variance compared to the variance of the PEEE approach)}
\end{table}

The simulation results revealed that both type B multiple imputation and the PEEE approach provided virtually unbiased estimates. However, type B multiple imputation based on 5 imputations provided estimates that exhibited a somewhat larger variability compared to the PEEE approach as a result of the finite number of imputations. With 20 imputations, the variability of the type B and PEEE approaches was similar. Standard error estimates were close to the corresponding Monte Carlo standard deviation of the estimates, and empirical coverage probabilities close to the nominal level in all cases. This provides numerical evidence for the validity of our closed-form variance estimator and our asymptotic normality result. 

The PEEE approach was substantially more computationally efficient compared to type B multiple imputation in terms of point estimate calculation. On average, the PEEE approach was about 6.4 times faster compared to type B multiple imputation based on 10 imputations, for simulation scenarios with 32.2\% missingness. For scenarios with 48.1\% missingness, the PEEE approach was 11 times faster compared to the type B multiple imputation approach with 10 imputations. This relative computational efficiency advantage of the PEEE approach did not seem to depend on sample size. In addition, we evaluated the computational time gains from using the proposed closed-form variance estimator for the PEEE approach instead of the bootstrap method. The corresponding results are presented in Table~\ref{sim_times}. The computation of the proposed closed-form variance estimator was about 70 to 90 times faster than the bootstrap method with 100 replications for the simulation scenarios with 32.2\% missingness. The corresponding figures for the scenarios with 48.1\% missingness ranged from to 52 to 62. The lower computational speed gains in scenarios with a larger missingness percent are attributed to the fact that fitting the multinomial logistic model for the incomplete variable to the complete cases requires more computation time compared to fitting the binary logit model during the main analysis. Therefore, estimation of the former model is faster in scenarios with a higher missingness rate and, thus, fewer complete cases. It must be noted that, in both sets of scenarios, the computational efficiency gains from our proposed closed-form variance estimator over the bootstrap method were more pronounced for larger sample sizes. However, the rate of increase in computational efficiency gain was declining with sample size. 

\begin{table}
\caption{Simulation results on computation times in seconds for the proposed closed-form variance estimator and the bootstrap method for variance estimation based on 100 replications.}
\label{sim_times}
\centering
\begin{tabular}{cccccc}
\toprule
&& \multicolumn{2}{c}{Proposed} & \multicolumn{2}{c}{Bootstrap}\\
\cmidrule(lr){3-4}\cmidrule(lr){5-6}
Missingness & $n$ & Mean & SD & Mean & SD \\
\hline
32.2\% & 1000 & 0.064 & 0.009 & 4.533 & 0.915 \\
& 2000 & 0.102 & 0.011 & 8.429 & 0.989 \\
& 3000 & 0.140 & 0.013 & 11.516 & 1.067 \\
& 4000 & 0.182 & 0.016 & 15.458 & 1.797 \\
& 5000 & 0.221 & 0.018 & 18.982 & 2.104 \\
& 6000 & 0.271 & 0.024 & 22.854 & 2.365 \\
& 7000 & 0.309 & 0.022 & 26.842 & 2.198 \\
& 8000 & 0.356 & 0.025 & 31.057 & 2.250 \\
& 9000 & 0.402 & 0.031 & 35.445 & 3.062 \\
& 10000 & 0.445 & 0.035 & 39.752 & 3.381 \\[1ex]
48.1\% & 1000 & 0.068 & 0.007 & 3.542 & 0.638 \\
& 2000 & 0.111 & 0.008 & 6.590 & 0.856 \\
& 3000 & 0.158 & 0.012 & 9.786 & 1.203 \\
& 4000 & 0.209 & 0.017 & 12.702 & 1.579 \\
& 5000 & 0.250 & 0.022 & 15.434 & 1.922 \\
& 6000 & 0.310 & 0.024 & 18.583 & 2.107 \\
& 7000 & 0.355 & 0.028 & 21.567 & 2.097 \\
& 8000 & 0.403 & 0.029 & 24.960 & 2.985 \\
& 9000 & 0.450 & 0.027 & 27.784 & 2.367 \\
& 10000 & 0.502 & 0.034 & 30.947 & 2.841 \\
\bottomrule
\end{tabular}
\\
\end{table}

To sum up, the simulation results indicate both the validity and the computational efficiency of the proposed closed-form variance estimator. The latter is particularly important in applications with larger datasets. 
A more extended evaluation of the performance of the proposed closed-form variance estimator under more pronounced model misspecification for the incomplete variable is provided in the simulation studies presented in Appendix C.

\section{Data Analysis}
\label{sec5}

The PEEE approach and our proposed variance estimator were used to analyze the data from the motivating TBI study. The study included electronic health records of 88,168 individuals with moderate/severe TBI who were hospitalized within 24 hours of their diagnosis. The data were obtained from the Indiana Registry of Traumatic Brain Injury, which has been described elsewhere \citep{Rahurkar17}. The scientific interest in this study was to evaluate predictors of rehabilitation within 3 months from hospitalization date and assess disparities in the receipt of rehabilitation services. The main covariates of interest were gender, age at hospitalization, being in the emergency room prior to the TBI diagnosis, and race (white, African American, and other race). Race was missing for 25.5\% of the patients, which limits the ability to detect health disparities in receipt of rehabilitation services post-injury. In this analysis, we considered the calendar year of admission as an auxiliary variable that is potentially associated with the probability of missingness. Given that one of the goals in this study was to use the fitted model to predict future cases of rehabilitation, it is not desirable to include calendar year as a covariate in the main analysis model. The results of the analysis based on the PEEE approach and the proposed closed-form variance estimator are presented in Table~\ref{TBI_res}.

\begin{table}
\caption{Predictors of rehabilitation services within three months of hospitalization in patients with TBI.}
\label{TBI_res}
\centering
\begin{tabular}{lccc}
\toprule
Covariate & OR & 95\% CI & $p$-value \\
\hline
Gender & & & \\
\ \ \textit{Female} & 1 & - & - \\
\ \ \textit{Male} & 1.029 &  (0.984, 1.076) & 0.206 \\
Age (10 years $\uparrow$) & 1.454 & (1.453, 1.456) & $<$0.001  \\
Race & & & \\
\ \ \textit{White} & 1 & - & - \\
\ \ \textit{Black or AA} & 0.940 & (0.875, 1.010) & 0.092 \\
\ \ \textit{Other} & 0.673 & (0.608, 0.745) & $<$0.001 \\
ER & 7.130 & (6.749, 7.534) & $<$0.001 \\
\bottomrule
\end{tabular}
\\
{OR: Odds ratio; AA: African American; ER: Admission to emergency room prior to TBI diagnosis}
\end{table}

Computation of point estimates based on the PEEE and standard errors using the proposed closed-form estimator required about 10 seconds. Variance estimation based on 100 bootstrap samples required about 10 minutes for the PEEE approach, and approximately 45 minutes for type B multiple imputation based on 10 imputations. This shows a significant computational efficiency advantage of the proposed closed-form variance estimator over the bootstrap method. Based on the PEEE approach, older patients, those with an emergency room admission prior to TBI diagnosis and white patients (compared to other non-whites and non-African-Americans), were more likely to receive rehabilitation services within 3 months from the hospitalization date. There was also a marginally non-significant difference between African Americans and whites with respect to the odds of rehabilitation services within 3 months from the hospitalization date ($p$-value=0.092). The analysis based on type B multiple imputation based on 10 imputations provided similar results to those obtained from the PEEE approach. The computation of our closed-form variance estimator in this data analysis using R is described in detail in Appendix D.

\section{Discussion}
\label{sec6}

In this paper we proposed closed-form variance estimators for the PEEE approach for parametric statistical analysis with missing data. This approach is in general more computationally-efficient compared to competing alternatives for missing data and, therefore, it is particularly useful for the analysis of larger datasets or when the main analysis requires computationally intensive methods. In this work we showed that the general PEEE estimator is consistent and asymptotically normal. Our key contribution was the derivation of closed-form variance estimators for the PEEE estimator which avoid the computational intensity of bootstrap methods for variance estimation. Our variance estimators where shown to be consistent even under uncongeniality between the main analysis and incomplete variable model as well as under misspecification of the latter model. A situation which leads to uncongeniality in practice is the inclusion of auxiliary variables in the incomplete variable model, in an effort to make the MAR assumption more plausible. Simulation studies showed a good performance of the estimators under both misspecification of the incomplete variable model and uncongeniality. Furthermore, the computation times for the proposed variance estimator were over 50 times faster compared to the bootstrap method for variance estimation in all scenarios. The computational efficiency of the PEEE and the proposed variance estimator was also illustrated in the application of the method to the motivating dataset of 88,168 patients with TBI. Calculation of the PEEE estimates and the proposed closed-form variance required 10 seconds on a regular personal computer, without utilizing parallel computing. The computation time for the variance based on 100 bootstrap samples required $\sim$10 minutes. The corresponding figure for the analysis based on type B multiple imputation with 10 imputations was $\sim$45 minutes, which indicates a substantial computational advantage of the PEEE approach with the proposed closed-form variance estimator in practice.

The PEEE approach was first introduced by \citet{Wang08} in a set of examples, as a computationally faster alternative to their EEE approach. However, \citet{Wang08} neither provided a variance estimator for the general PEEE approach nor showed the consistency and asymptotic normality of the PEEE estimator. Moreover, they did not provide details on the calculation of the PEEE estimates for general missing data problems in practice. In this paper, we have addressed all these important gaps. We provided a closed-form variance estimator which is consistent even under uncongeniality and  misspecification of the model for the incomplete variable, unlike Rubin's variance estimator for type A multiple imputation. Our closed-form variance estimator is particularly useful in applications with large data sets as it can considerably reduce the computation time compared to bootstrap methods for variance estimation, as illustrated in our simulation study and data analysis. Furthermore, we rigorously established the consistency and asymptotic normality of the PEEE estimator using modern empirical process theory \citep{van96, Kosorok08}. This allowed us to assume that the estimating function $\psi_{\theta}$ for the main analysis model is just Lipschitz continuous in $\theta$ in a neighborhood of the true $\theta_0$. This is satisfied if $\psi_{\theta}$ is (one time) differentiable with respect to $\theta$ in a neighborhood of $\theta_0$, and if the (first) derivative has a bounded second moment. In contrast, using more traditional asymptotic theory tools would require the stronger assumption that $\psi_{\theta}$ is twice continuously differentiable in a neighborhood of the true $\theta_0$, that the first derivative has a bounded second moment, and that the second derivative has a bounded expectation \citep{van96}. Last but not least, we provided details on how to compute the PEEE estimator using standard software in Appendix B.

The PEEE approach is related to other parametric methods for analyzing incomplete data. First, it is a two-stage (and computationally more efficient) version of the EEE approach \citep{Wang08}. Second, the parametric type B multiple imputation \citep{Wang98} is equivalent to the Monte Carlo PEEE approach which is used in cases where the conditional expectation $E[\psi_{\theta}(X^o,X^m)|X^o=x^o,A=a;\gamma_0]$ cannot be specified in closed-form. Third, the Monte Carlo PEEE approach is also a version of parametric fractional imputation. More precisely, the missing values are simulated from $F_{X^m|X^o,A}(x|X^o_i,A_i;\hat{\gamma}_n)$ (special choice for the \textit{proposal distribution} \citep{Kim11}) and the fractional weights are equal to $1/S$, where $S$ is the number of imputations. Finally, the PEEE approach reduces to single regression imputation in the special case where the incomplete variable is linear in the estimating function. Regardless of its connection with other parametric methods for missing data, the PEEE can be more computationally efficient and (slightly) more statistically efficient than parametric multiple imputation approaches. Moreover, unlike single regression imputation, the PEEE estimator is consistent even if the incomplete variable is not linear in the estimating function. These advantages of the PEEE approach make it a useful general tool for modern applications with large incomplete datasets.

A limitation of the PEEE approach is that, similarly to parametric multiple imputation, it assumes a parametric model for the incomplete variable. The reason we relied on parametric models is due to their practical utility. If the model for the incomplete variable is misspecified, then the PEEE estimator converges to a point $\theta^*$ which is not equal to the true parameter value $\theta_0$ (even though the proposed variance estimator is consistent regardless of the correct specification of the latter model). The simulation studies under various degrees of model misspecification (see Appendix C) provide numerical evidence that the bias of the PEEE estimator is virtually identical to that of type B multiple imputation. When $\hat{\gamma}_n$ is a maximum likelihood estimator, $\gamma^*$ is the minimizer of the Kullback--Leibler divergence between the assumed model and the true model for the incomplete variable. In such cases, the limit $\theta^*$ is the solution to the (true) expected estimating equations under the \textit{closest} (with respect to the Kullback--Leibler divergence) model to the true one. Regardless of the estimation method used, a remedy to reduce the impact of model misspecification is to consider flexible parametric models including, for example, regression splines for the continuous variables. Numerical evidence for the effectiveness of the latter approach for dealing with misspecification due to non-linearity is provided by a series of simulation experiments that are reported in Appendix C. For a general nonparametric kernel-based method for missing data we refer the reader to the excellent work by \citet{Zhou08}. 

We can see several important topics for future research. First, considering nonparametric machine learning methods for estimating $E[\psi_{\theta}(X^o,X^m)|X^o=x^o,A=a]$ is particularly useful in order to avoid the bias in the point estimate due to general model misspecification. In addition, this would be interesting from a theoretical standpoint. Second, incorporating misclassification and measurement error as in the EEE approach by \citep{Wang08} and providing a closed-form variance estimator is crucial since such complications arise frequenty in medical research. Third, extending the general PEEE approach to situations where the main analysis model is semiparametric would also be interesting and useful from a practical standpoint. This would have important applications in survival analysis with large, incomplete datasets.

\section*{Acknowledgments}

This project was supported, in part, with support from the National Institute of Allergy and Infectious Diseases, grant numbers R01AI140854 and R21AI145662, and the Indiana Spinal Cord \& Brain Injury Research Fund from the Indiana State Department of Health. Its contents are solely the responsibility of the authors and do not necessarily represent the official views of the Indiana State Department of Health. The authors further acknowledge the Regenstrief Institute for facilitating access to the data, especially Ashley Wiensch, MPH.
{\it Conflict of Interest}: None declared.

\bibliographystyle{chicago}
\bibliography{refs}

\newpage
\begin{appendices}

\section*{Appendix A: Proofs}
\renewcommand{\thesubsection}{A.\arabic{subsection}}
\label{s:intro}
The proofs of the asymptotic properties will rely on modern empirical process theory \citep{van96,Kosorok08}. We will use the standard empirical process notation $Pf=E[f(D)]$ for a measurable function $f:\mathcal{D}\mapsto\mathbb{R}$, where $\mathcal{D}$ is the sample space, and $\mathbb{P}_nf\equiv n^{-1}\sum_{i=1}^nf(D_i)$.

The asymptotic properties of the PEEE estimator with rely on the following assumptions and regularity conditions. First we assume that the model is identifiable. More precisely, we assume that the fixed function $P\psi_{\theta}\equiv \Psi(\theta)$ satisfies $\|\Psi(\theta_0)\|=0$, and for any sequence $\{\theta_n\}_{n\geq 1}\subset\Theta$, $\|\Psi(\theta_n)\|\rightarrow 0$ implies that $\|\theta_n-\theta\|\rightarrow 0$. Additionally, we assume MAR conditional on both the observed and the auxiliary variables, that is:
\[
P(R=1|X^m,X^o,A)=P(R=1|X^o,A).
\]
We also assume the following regularity conditions.
\begin{itemize}
\item[C1.] The true parameter $\theta_0$ lies in the interior of a convex and bounded set $\Theta\subset\mathbb{R}^d$.
\item[C2.] The function $\theta \mapsto \psi_{\theta}(x)$ is continuous for all $x$ in the sample space. Moreover, for any $\theta_1, \theta_2\in\Theta_{\delta}=\{\theta\in\Theta:\|\theta-\theta_0\|<\delta\}$ for some $\delta>0$ we have
\[
\|\psi_{\theta_1}(x)-\psi_{\theta_2}(x)\|\leq \dot{\psi}(x)\|\theta_1 - \theta_2\|
\]
for some (measurable) function $\dot{\psi}$ which satisfies $P\dot{\psi}^2<\infty$.
\item[C3.] The PEEE and MCPEEE estimators satisfy  $\|\tilde{\Psi}_n(\hat{\theta}_n;\hat{\gamma}_n)\|=o_p(n^{-1/2})$ and $\|\tilde{\Psi}_{n,S}(\hat{\theta}_{n,S};\hat{\gamma}_n)\|=o_p(n^{-1/2})$.
\item[C4.] The fixed function $\theta\mapsto\Psi(\theta)$ is continuously differentiable at $\theta_0$ with a continuously invertible derivative $\dot{\Psi}_{\theta_0}$.
\item[C5.] The model for the incomplete variable $F_{X^m|X^o,A}(x|x^o,a;\gamma_0)$ is correctly specified. Also, the parameter $\gamma_0$ lies in the interior of the corresponding parameter space $\Gamma$ and $\Gamma$ is a convex and bounded subset of $\mathbb{R}^p$. In addition, $(X,A)$ is bounded in the sense that there exists a constant $C_1\in(0,\infty)$ such that $P(\|X\|\vee\|A\|\leq C_1)=1$.
\item[C6.]The estimator for $\gamma_0$ satisfies $\|\hat{\gamma}_n-\gamma_0\|=o_p(1)$ and is an asymptotically linear estimator with $\sqrt{n}(\hat{\gamma}_n-\gamma_0)=n^{-1/2}\sum_{i=1}^n\omega_i+o_p(1)$, where $E\omega=0$ and $E\|\omega\|^2<\infty$. Additionally, the empirical versions of the influence functions $\hat{\omega}_i$, $i=1,\ldots,n$, satisfy $n^{-1}\sum_{i=1}^n\|\hat{\omega}_i-\omega_i\|^2=o_p(1)$.
\item[C7.] $E[\psi_{\theta}(X)|X^o=x^o,A=a;\gamma]$ is continuously differentiable with respect to $\gamma$ on compacts, for all $x^o$ and $a$ in the sample space.
\item[C7$^{\prime}$.] $\psi_{\theta}(x^o,\tilde{x}^m(u,\gamma))=\psi_{\theta}(x^o,F_{X^m|X^o,A}^{-1}(u|X^o=x^o,A=a;\gamma))$ is continuously differentiable with respect to $\gamma$ on compacts, for all $x^o$ and $a$ in the sample space.
\end{itemize}

Conditions C1--C4 are standard conditions that ensure the consistency and asymptotic normality of $Z$-estimators \citep{Kosorok08}. These conditions are mild and will be typically satisfied for the complete-data analysis.  Conditions C5--C7$^{\prime}$ are required in order to handle missingness.  Condition C5 will be automatically satisfied if $\hat{\gamma}_n$ is a maximum likelihood estimated under a correctly specified model. If the model is misspecified, then condition C5 still holds but with $\gamma_0$ being replaced by $\gamma^*$, which is the parameter that minimizes the Kullback--Leibler divergence between the assumed model and the true model. If $\gamma_0$ is estimated via least squares under a correctly specified model, then condition C6 is still satisfied. 
There are of course many other examples of estimators that satisfy condition C6. Condition C7 assumes that the conditional expectation in the estimating function is sufficiently smooth in both $\gamma$. The latter is a realistic assumption with parametric models in general.

\subsection{Proof of Theorem 1}
To prove consistency we will show the consistency conditions for Z-estimators \citep{Kosorok08}. We first prove the consistency of the PEEE estimator. Initially, we have that
\begin{eqnarray*}
\sup_{\theta\in\Theta}\|\mathbb{P}_n\tilde{\psi}_{\theta,\hat{\gamma}_n}-P\psi_{\theta}\|&\leq&\sup_{\theta\in\Theta}\|\mathbb{P}_n(\tilde{\psi}_{\theta,\hat{\gamma}_n}-\tilde{\psi}_{\theta,\gamma_0})\| \\
&&+\sup_{\theta\in\Theta}\|(\mathbb{P}_n-P)\tilde{\psi}_{\theta,\gamma_0}\| \\
&& +\sup_{\theta\in\Theta}\|P\tilde{\psi}_{\theta,\gamma_0}-P\psi_{\theta}\|\\
&\equiv& A^{\prime}_n+B^{\prime}_n+C^{\prime}
\end{eqnarray*}
Letting $W=(X^o,A)$ for notational simplicity, it is easy to see that the first term above is
\begin{eqnarray*}
A^{\prime}_n&\leq&\sup_{\theta\in\Theta}\left\|\frac{1}{n}\sum_{i=1}^n\{E[\psi_{\theta}(X^o,X^m)|W_i;\hat{\gamma}_n]-E[\psi_{\theta}(X^o,X^m)|W_i;\gamma_0]\}\right\|.
\end{eqnarray*}
Condition C7 implies that for any $\gamma_1,\gamma_2\in\Gamma$ and any $w$ in the corresponding sample space $\mathcal{W}$ we have that
\[
\left\|E[\psi_{\theta}(X)|W=w;\gamma_1]-E[\psi_{\theta}(X)|W=w;\gamma_2]\right\|\leq K(\theta,w)\|\gamma_1-\gamma_2\|
\]
with $C_2\equiv\sup_{\theta,w}K(\theta,w)<\infty$. Thus, by condition C6, it follows that,
\[
A^{\prime}_n\leq C_2\|\hat{\gamma}_n-\gamma_0\|=o_p(1).
\]
For the second term $B^{\prime}_n$ consider the class of functions $\mathcal{F}_{1}=\{\tilde{\psi}_{\theta,\gamma_0}:\theta\in\Theta\}$, where
\[
\tilde{\psi}_{\theta,\gamma_0}(D)=R\psi_{\theta}(X^o,X^m)+(1-R)E[\psi_{\theta}(X^o,X^m)|W;\gamma_0].
\]
Note that the fixed class $\{\theta:\theta\in\Theta\subset\mathbb{R}^d\}$ is trivially Donsker, and, thus, also Glivenko--Cantelli. Now, the classes $\{\psi_{\theta}:\theta\in\Theta\}$ and $\{m_{\theta,\gamma_0}:\theta\in\Theta\}$, with $m_{\theta,\gamma_0}(W)=E[\psi_{\theta}(X)|W;\gamma_0]$, are both $P$-Glivenko--Cantelli by condition C2 and Corollary 9.27 in \citet{Kosorok08}. Therefore, the class $\mathcal{F}_1$ is $P$-Glivenko--Cantelli as it is formed by the sum of two $P$-Glivenko--Cantelli classes which are multiplied by random variables with finite second moments. Thus, $B^{\prime}_n\overset{as*}\rightarrow 0$. Finally, after some algebra it can be shown that $C^{\prime}=0$ by condition C5 and, therefore, $\sup_{\theta\in\Theta}\|\mathbb{P}_n\tilde{\psi}_{\theta,\hat{\gamma}_n}-P\psi_{\theta}\|=o_p(1)$. This fact along with the identifiability assumption lead to the conclusion that
\[
\|\hat{\theta}_n-\theta_0\|\overset{p}\rightarrow 0,
\]
as $n\rightarrow\infty$. 

Next, we show the consistency of the MCPEEE estimator. First, we have that
\begin{eqnarray*}
\sup_{\theta\in\Theta}\|\mathbb{P}_n\tilde{\psi}_{\theta,\hat{\gamma}_n,S}-P\psi_{\theta}\|&\leq&\sup_{\theta\in\Theta}\|\mathbb{P}_n(\tilde{\psi}_{\theta,\hat{\gamma}_n,S}-\tilde{\psi}_{\theta,\gamma_0,S})\| \\
&&+\sup_{\theta\in\Theta}\|(\mathbb{P}_n-P)\tilde{\psi}_{\theta,\gamma_0,S}\| \\
&& +\sup_{\theta\in\Theta}\|P\tilde{\psi}_{\theta,\gamma_0,S}-P\tilde{\psi}_{\theta,\gamma_0}\| \\
&& +\sup_{\theta\in\Theta}\|P\tilde{\psi}_{\theta,\gamma_0}-P\psi_{\theta}\|.
\end{eqnarray*}
Using similar arguments to those used for the consistency of the PEEE estimator it can be shown that $\sup_{\theta\in\Theta}\|\mathbb{P}_n(\tilde{\psi}_{\theta,\hat{\gamma}_n,S}-\tilde{\psi}_{\theta,\gamma_0,S})\|=o_p(1)$ (by conditions C6 and C7$^{\prime}$), $\sup_{\theta\in\Theta}\|(\mathbb{P}_n-P)\tilde{\psi}_{\theta,\gamma_0,S}\|=o_{as*}(1)$, and $\sup_{\theta\in\Theta}\|P\tilde{\psi}_{\theta,\gamma_0}-P\psi_{\theta}\|=0$. For the remaining term we have that
\[
\sup_{\theta\in\Theta}\|P\tilde{\psi}_{\theta,\gamma_0,S}-P\tilde{\psi}_{\theta,\gamma_0}\|\leq\sup_{\theta\in\bar{\Theta}}\left\|E[\psi_{\theta}(X^o,\tilde{X}^m(U,\gamma_0))]-E[\psi_{\theta}(X^o,X^m)]\right\|=0,
\]
where $\bar{\Theta}$ is the closure of the set $\Theta$, by the i.i.d. assumption and the fact that $(X^o,\tilde{X}^m(U,\gamma_0))$ and $(X^o,X^m)$ follow the same distribution. The latter statement follows from the fact that $\tilde{X}^m(U,\gamma_0)$ and $X^m$ follow the same conditional distribution given $(X^o,A)$ by condition C5 and, thus, $(\tilde{X}^m(U,\gamma_0),X^o,A)$ and $(X^m,X^o,A)$ follow the same distribution as well. Therefore, \[
\sup_{\theta\in\Theta}\|\mathbb{P}_n\tilde{\psi}_{\theta,\hat{\gamma}_n,S}-P\psi_{\theta}\|=o_p(1),
\]
and thus, by the identifiability condition,
\[
\|\hat{\theta}_{n,S}-\theta_0\|\overset{p}{\rightarrow} 0
\]
as $n\rightarrow\infty$.

\subsection{Proof of Theorem 2}
By condition C3 we have that
\begin{eqnarray}
o_p(1)&=&\sqrt{n}\tilde{\Psi}_n(\hat{\theta}_n;\hat{\gamma}_n) \nonumber \\
&=& \sqrt{n}\left[\tilde{\Psi}_n(\hat{\theta}_n;\hat{\gamma}_n)-\tilde{\Psi}_n(\hat{\theta}_n;\gamma_0)\right]+\sqrt{n}\tilde{\Psi}_n(\hat{\theta}_n;\gamma_0) \nonumber \\
&\equiv& A_n^{\prime\prime}+B_n^{\prime\prime} \label{psi0}
\end{eqnarray}
Letting $m_{\theta,\gamma}(W)\equiv E[\psi_{\theta}(X^o,X^m)|W;\gamma]$ and after some algebra we have that
\begin{eqnarray*}
A_n^{\prime\prime}&=&\sqrt{n}\mathbb{P}_n(1-R)\left[m_{\hat{\theta}_n,\hat{\gamma}_n}(W)-m_{\hat{\theta}_n,\gamma_0}(W)\right] \\
&=&\sqrt{n}(\mathbb{P}_n-P)(1-R)\left[m_{\hat{\theta}_n,\hat{\gamma}_n}(W)-m_{\theta_0,\gamma_0}(W)\right] \\
&&-\sqrt{n}(\mathbb{P}_n-P)(1-R)\left[m_{\hat{\theta}_n,\gamma_0}(W)-m_{\theta_0,\gamma_0}(W)\right] \\
&&+\sqrt{n}P(1-R)\left[m_{\hat{\theta}_n,\hat{\gamma}_n}(W)-m_{\hat{\theta}_n,\gamma_0}(W)\right] \\
&\equiv&A_{n,1}^{\prime\prime}-A_{n,2}^{\prime\prime}+A_{n,3}^{\prime\prime}
\end{eqnarray*}
Next, consider the class of functions $\mathcal{F}_2=\{m_{\theta,\gamma}-m_{\theta_0,\gamma_0}:\theta\in\Theta_{\delta},\gamma\in\Gamma_{\delta}\}$, with $\Gamma_{\delta}=\{\gamma\in\Gamma:\|\gamma-\gamma_0\|<\delta\}$, for some $\delta>0$. Given that the fixed classes $\Theta_{\delta}$ and $\Gamma_{\delta}$ are trivially $P$-Donsker, and by conditions C2 and C7 and Corollary 9.32 in \citet{Kosorok08}, the class $\mathcal{F}_2$ is also $P$-Donsker. Next, we have that 
\begin{eqnarray*}
P(m_{\theta,\gamma}-m_{\theta_0,\gamma_0})^2&\leq&P(m_{\theta,\gamma}-m_{\theta_0,\gamma})^2+P(m_{\theta_0,\gamma}-m_{\theta_0,\gamma_0})^2 \\
&&+2P(\|m_{\theta,\gamma}-m_{\theta_0,\gamma}\|\|m_{\theta_0,\gamma}-m_{\theta_0,\gamma_0}\|) \\
&\leq&\|\theta-\theta_0\|^2E\left(\left\{E\left[\dot{\psi}(X^o,X^m)|W,\gamma\right]\right\}^2\right)+\|\gamma-\gamma_0\|^2C_2^2 \\
&& +2\|\theta-\theta_0\|\|\gamma-\gamma_0\|\max(1,P\dot{\psi}^2)C_2 \\
&\leq&\|\theta-\theta_0\|^2P\dot{\psi}^2+\|\gamma-\gamma_0\|^2C_2^2+2\|\theta-\theta_0\|\|\gamma-\gamma_0\|\max(1,P\dot{\psi}^2)C_2
\end{eqnarray*}
where the first inequality follows from the Cauchy-Schwartz inequality, the second from conditions C2 and C7, the third from Jensen's inequality and the law of total expectation, and $C_2$ is the Lipschitz constant for the map $\gamma\mapsto E[\psi_{\theta_0}(X^o,X^m)|W,\gamma]$, in light of condition C7. Thus, by condition C2 we have that $P(m_{\theta,\gamma}-m_{\theta_0,\gamma_0})^2\rightarrow 0$ as $\|\theta-\theta_0\|\rightarrow 0$ and $\|\gamma-\gamma_0\|\rightarrow 0$. These facts along with the consistency in probability of $\hat{\theta}_n$ and $\hat{\gamma}_n$ and arguments similar to those used in the proof of Lemma 3.3.5 in \citet{van96} imply that $A_{n,1}''=o_p(1)$. Using the same arguments one can also show that $A_{n,2}''=o_p(1)$.

Next, condition C7 and a Taylor expansion at $\gamma_0$ lead to
\begin{eqnarray*}
A_{n,3}^{\prime\prime}&=&\{P[(1-R)\dot{m}_{\hat{\theta}_n,\gamma_0}(W)]\}\sqrt{n}(\hat{\gamma}_n-\gamma_0)+o_p(1) 
\\
&=&\{P[(1-R)\dot{m}_{\hat{\theta}_n,\gamma_0}(W)]-P[(1-R)\dot{m}_{\theta_0,\gamma_0}(W)]\}\sqrt{n}(\hat{\gamma}_n-\gamma_0) \\
&&+\{P[(1-R)\dot{m}_{\theta_0,\gamma_0}(W)]\}\sqrt{n}(\hat{\gamma}_n-\gamma_0)+o_p(1),
\end{eqnarray*}
where
\[
\dot{m}_{\hat{\theta}_n,\gamma_0}(W)=\frac{\partial}{\partial\gamma}E[\psi_{\hat{\theta}_n}(X)|W,\gamma]\big|_{\gamma=\gamma_0}.
\]
By the consistency of $\hat{\theta}_n$, the continuous mapping theorem, and the fact that $\sqrt{n}(\hat{\gamma}_n-\gamma_0)=O_P(1)$ in light of condition C6, it follows that
\[
\{P[(1-R)\dot{m}_{\hat{\theta}_n,\gamma_0}(W)]-P[(1-R)\dot{m}_{\theta_0,\gamma_0}(W)]\}\sqrt{n}(\hat{\gamma}_n-\gamma_0)=o_p(1).
\]
Thus, by condition C6, $A_{n,3}^{\prime\prime}$ can be expressed as an asymptotically linear term and, therefore,
\begin{eqnarray}
A_{n}^{\prime\prime}&=&\{P[(1-R)\dot{m}_{\theta_0,\gamma_0}(W)]\}\sqrt{n}\mathbb{P}_n\omega+o_p(1) \nonumber \\
&\equiv& G_{\theta_0,\gamma_0}\sqrt{n}\mathbb{P}_n\omega+o_p(1). \label{A_n}
\end{eqnarray}

Next, from \eqref{psi0} we have that
\begin{equation}
-A_n^{\prime\prime}-\sqrt{n}[\tilde{\Psi}_n(\hat{\theta}_n;\gamma_0)-\Psi(\hat{\theta}_n)]+o_p(1)=\sqrt{n}[\Psi(\hat{\theta}_n)-\Psi(\theta_0)] \label{psi0_1}.
\end{equation}
Note that, by condition C5, we have that $\Psi(\theta)=P\tilde{\psi}_{\theta,\gamma_0}$ for any $\theta\in\Theta$. Next, consider the class of functions
\[
\mathcal{F}_3=\{\tilde{\psi}_{\theta,\gamma_0}-\tilde{\psi}_{\theta_0,\gamma_0}:\theta\in\Theta_{\delta}\}.
\]
Note that
\[
\tilde{\psi}_{\theta,\gamma_0}(D)-\tilde{\psi}_{\theta_0,\gamma_0}(D)=R[\psi_{\theta}(X^o,X^m)-\psi_{\theta_0}(X^o,X^m)] +(1-R)[m_{\theta,\gamma_0}(W)-m_{\theta_0,\gamma_0}(W)].
\]
The class $\{\psi_{\theta}-\psi_{\theta_0}:\theta\in\Theta_{\delta}\}$ is $P$-Donsker by condition C2 and corollary 9.32 in \cite{Kosorok08}. Also, the class $\{m_{\theta,\gamma_0}-m_{\theta_0,\gamma_0}:\theta\in\Theta_{\delta}\}$ is $P$-Donsker as it is a subset of the $P$-Donsker class $\mathcal{F}_2$ and, thus, the class $\mathcal{F}_3$ is $P$-Donsker as well as it is formed by the sum of two $P$-Donsker classes which are multiplied by random variables with finite second moments. Moreover, by calculations and arguments similar to those used for $P(m_{\theta,\gamma}-m_{\theta_0,\gamma_0})^2$ above, we have that $P(\tilde{\psi}_{\theta,\gamma_0}-\tilde{\psi}_{\theta_0,\gamma_0})^2\rightarrow 0$ as $\|\theta-\theta_0\|\rightarrow 0$. Therefore, by the consistency of $\hat{\theta}_n$ and arguments similar to those used in the proof of Lemma 3.3.5 in \citet{van96} it follows that
\[
\sqrt{n}(\mathbb{P}_n-P)(\tilde{\psi}_{\hat{\theta}_n,\gamma_0}-\tilde{\psi}_{\theta_0,\gamma_0})=o_p(1).
\]
Thus, \eqref{psi0_1} can be rewritten as
\begin{equation}
-A_n^{\prime\prime}-\sqrt{n}(\mathbb{P}_n-P)\tilde{\psi}_{\theta_0,\gamma_0}+o_p(1)=\sqrt{n}[\Psi(\hat{\theta}_n)-\Psi(\theta_0)]. \label{psi0_2}
\end{equation}
Next, by the differentiability of $\Psi$ (condition C4), and condition C5 which implies that $P\tilde{\psi}_{\theta_0,\gamma_0}=\Psi(\theta_0)=0$, \eqref{psi0_2} becomes
\[
\sqrt{n}\dot{\Psi}_{\theta_0}(\hat{\theta}_n-\theta_0)=-A_n^{\prime\prime}-\sqrt{n}\mathbb{P}_n\tilde{\psi}_{\theta_0,\gamma_0}+o_p(1+\sqrt{n}\|\theta-\theta_0\|).
\]
Next we will show that $o_p(1+\sqrt{n}\|\theta-\theta_0\|)=o_p(1)$. By the invertability of $\dot{\Psi}_{\theta_0}$ (condition C4) it follows that $\|\dot{\Psi}_{\theta_0}(\theta-\theta_0)\|\geq c\|(\theta-\theta_0)\|$ for some $c>0$ and all $\theta$ and $\theta_0$. Thus, by the differentiability of $\Psi$ (condition C4), $\|\Psi(\theta)-\Psi(\theta_0)\|\geq c\|\theta-\theta_0\|+o(\|\theta-\theta_0\|)$. This fact, along with \eqref{psi0_2}, and the central limit theorem applied to $A_n^{\prime\prime}$ and $\sqrt{n}(\mathbb{P}_n-P)\tilde{\psi}_{\theta_0,\gamma_0}$, lead to
\[
\sqrt{n}\|\hat{\theta}_n-\theta_0\|(c+o_p(1))\leq O_p(1)+O_p(1)+o_p(1)
\]
and, therefore, $\sqrt{n}\|\hat{\theta}_n-\theta_0\|=O_p(1)$.
Thus $o_p(1+\sqrt{n}\|\theta-\theta_0\|)=o_p(1)$ and, therefore, by condition C4 and the asymptotic linear representation of $A_n^{\prime\prime}$ it follows that
\[
\sqrt{n}(\hat{\theta}_n-\theta_0)=-\dot{\Psi}_{\theta_0}^{-1}\sqrt{n}\mathbb{P}_n(\tilde{\psi}_{\theta_0,\gamma_0}+G_{\theta_0,\gamma_0}\omega)+o_p(1).
\]
Therefore, by the fact that $P(\tilde{\psi}_{\theta_0,\gamma_0}+G_{\theta_0,\gamma_0}\omega)=0$ by conditions C5 and C6, along with the fact that $P\|\tilde{\psi}_{\theta_0,\gamma_0}+G_{\theta_0,\gamma_0}\omega\|^2<\infty$ by conditions C2, C5 and C6, and the central limit theorem it follows that
\[
\sqrt{n}(\hat{\theta}_n-\theta_0)\overset{d}\rightarrow N(0,\Omega),
\]
where
\[
\Omega=\left(\dot{\Psi}_{\theta_0}^{-1}\right)P\left(\tilde{\psi}_{\theta_0,\gamma_0}+G_{\theta_0,\gamma_0}\omega\right)^{\otimes 2}\left(\dot{\Psi}_{\theta_0}^{-1}\right),
\]
with $B^{\otimes 2}=BB^T$ for a matrix $B$.

The consistency of the variance estimator $\hat{\Omega}_n$ follows from the next arguments. First, by conditions C5, C6, the consistency of $\hat{\theta}_n$ and the extended continuous mapping theorem, theorem 7.24 in \cite{Kosorok08}, it follows that $\dot{\Psi}_{n,\hat{\theta}_n}(\hat{\gamma}_n)\overset{p}\rightarrow \dot{\Psi}_{\theta_0}$. The same arguments can be used to argue that
\[
\mathbb{P}_n\left(\tilde{\psi}_{\hat{\theta}_n,\hat{\gamma}_n}+\hat{G}_{\hat{\theta}_n,\hat{\gamma}_n}\hat{\omega}\right)^{\otimes 2}\overset{p}\rightarrow P\left(\tilde{\psi}_{\theta_0,\gamma_0}+G_{\theta_0,\gamma_0}\omega\right)^{\otimes 2}.
\]
Finally, the continuous mapping theorem implies that $\hat{\Omega}_n\overset{p}\rightarrow \Omega$.

\subsection{Proof of Theorem 3}
The proof of Theorem 3 follows from similar arguments to those used in the proof of Theorem 2. Here, we consider the alternative to $\mathcal{F}_2$ and $\mathcal{F}_3$ classes of functions 
\[
\mathcal{F}_2^{\prime}=\{m_{\theta,\gamma,S}-m_{\theta_0,\gamma_0,S}:\theta\in\Theta_{\delta},\gamma\in\Gamma_{\delta}\},
\]
where 
\[
m_{\theta,\gamma,S}(W)=\frac{1}{S}\sum_{j=1}^S\psi_{\theta}(X^o,F_{X^m|X^o,A}^{-1}(U_j|X^o,A;\gamma)),
\]
and 
\[
\mathcal{F}_3^{\prime}=\{\tilde{\psi}_{\theta,\gamma_0,S}-\tilde{\psi}_{\theta_0,\gamma_0,S}:\theta\in\Theta_{\delta}\}.
\]
Both $\mathcal{F}_2^{\prime}$ and $\mathcal{F}_3^{\prime}$ are $P$-Donsker for any finite $S$ by conditions C2 and C7$^{\prime}$, and Corollary 9.32 in \citet{Kosorok08}. In addition, $P\tilde{\psi}_{\theta,\gamma_0,S}=\Psi(\theta)$ for any $\theta\in\Theta$ and $S$ by the i.i.d. assumption, condition C5, and the fact that $(\tilde{X}^m(U,\gamma_0),X^o,A,R)$ and $(X^m,X^o,A,R)$ follow the same distribution. The latter is a consequence of the fact that $(\tilde{X}^m(U,\gamma_0),X^o,A)$ and $(X^m,X^o,A)$ follow the same distribution, as argued in the proof of Theorem 1, and the MAR assumption. These facts and similar arguments to those used in the proof of Theorem 2 lead to
\[
\sqrt{n}(\hat{\theta}_{n,S}-\theta_0)=-\dot{\Psi}_{\theta_0}^{-1}\sqrt{n}\mathbb{P}_n(\tilde{\psi}_{\theta_0,\gamma_0,S}+H_{\theta_0,\gamma_0}\omega)+o_p(1),
\]
where
\[
H_{\theta_0,\gamma_0}=E\left[(1-R)\frac{\partial}{\partial\gamma}\psi_{\theta}(X^o,F_{X^m|X^o,A}^{-1}(U|X^o,A;\gamma))|_{\gamma=\gamma_0}\right]
\]
Therefore, by the fact that $P(\tilde{\psi}_{\theta_0,\gamma_0,S}+H_{\theta_0,\gamma_0}\omega)=0$ by conditions C5 and C6, along with the fact that $P\|\tilde{\psi}_{\theta_0,\gamma_0,S}+H_{\theta_0,\gamma_0}\omega\|^2<\infty$ by conditions C2, C5 and C6, and the central limit theorem it follows that
\[
\sqrt{n}(\hat{\theta}_{n,S}-\theta_0)\overset{d}\rightarrow N(0,\Omega_S),
\]
with
\[
\Omega_S=\left(\dot{\Psi}_{\theta_0}^{-1}\right)P\left(\tilde{\psi}_{\theta_0,\gamma_0,S}+H_{\theta_0,\gamma_0}\omega\right)^{\otimes 2}\left(\dot{\Psi}_{\theta_0}^{-1}\right).
\]
The consistency of the variance estimator $\hat{\Omega}_{n,S}$ follows from similar arguments to those used in the proof of Theorem 2.

\section*{Appendix B: Computation of the PEEE estimator using off-the-shelf software}
\renewcommand{\thesubsection}{B.\arabic{subsection}}
\setcounter{subsection}{0}
The PEEE estimator can be easily obtained using any standard software that allows for case weights using a simple data augmentation approach. Now consider three typical cases: (i) $\psi_{\theta}(X^o,X^m)$ is linear in $X^m$, (ii) $\psi_{\theta}(X^o,X^m)$ is not linear in $X^m$ but $X^m$ is a discrete variable that takes its values in the finite set $\{x_1,\ldots,x_K\}$, and (iii) $\psi_{\theta}(X^o,X^m)$ is neither linear in $X^m$ nor $X^m$ takes its values in a finite set, and $E[\psi_{\theta}(X^o,X^m)|X^o=x^o,A=a;\gamma_0]$ cannot be specified in closed form. Importantly, for cases (i) and (ii) the PEEE approach avoids the additional variability due to the finite number of stochastic imputations in both Type A and Type B multiple imputation. Also, the PEEE approach is more computationally efficient as it can be implemented using a considerably smaller pseudo-complete dataset compared to the type B multiple imputation approach. For case (iii), one can utilize Monte Carlo methods to estimate the conditional expectation of the incomplete variable given the fully observed variables. In this case, the PEEE approach is equivalent to type B multiple imputation. However, we show how to compute the PEEE estimator using off-the-shelf software by only utilizing a dataset of only $n+m(S-1)$ observations, where $m$ is the number of incomplete observations and $S$ is the chosen number of imputations, instead of a dataset of size $nS$ for the type B multiple imputation approach.

For simplicity of presentation, the examples provided in the following sections involve linear, in the covariates, models for $E[\psi_{\theta}(X^o,X^m)|X^o=x^o,A=a;\gamma_0]$. However, the PEEE approach is applicable with any model that involves parametric nonlinear terms or with flexible models with a finite number of parameters, such as regression splines. In addition, we used the more compact notation $W\equiv(X^o,A)$ for the fully observed variables.

\subsection{Incomplete Variable Linear in $\psi_{\theta}(X^o,X^m)$}

For the case where $\psi_{\theta}(X_i)$ is linear in the incomplete variable $X_i^m$, it is easy to see that, for the missing cases,
\[
E[\psi_{\theta}(X^o,X^m)|W_i;\gamma_0]=\psi_{\theta}\left(X_i^o,E(X^m|W_i;\gamma_0)\right),
\]
In this case, the analysis can be performed by using an estimate $\hat{\gamma}_n$ based on the complete cases, and replacing the missing values by their conditional expectation under the assumed parametric model, with $\gamma_0$ being replaced by $\hat{\gamma}_n$. Therefore, estimation of $\theta_0$ can be performed using the PEEE
\[
\tilde{\Psi}_n(\theta;\hat{\gamma}_n)=\frac{1}{n}\sum_{i=1}^n \left[R_i\psi_{\theta}(X_i^o,X_i^m)+(1-R_i)\psi_{\theta}\left(X_i^o,E(X^m|W_i;\hat{\gamma}_n)\right)\right].
\]
In practice, estimation of $\theta_0$ can be performed using any software package, by replacing the missing values with $\psi_{\theta}\left(X_i^o,E(X^m|W_i;\hat{\gamma}_n)\right)$. The required data setup in this case is presented in Table~\ref{exlin}.

\begin{table}
\centering
\caption{Example of the dataset up for two hypothetical individuals $i$ and $i+1$ in the case of an incomplete variable $X^m$ which is linear in $\psi_{\theta}(X^o,X^m)$. ($R$: missingness indicator with $R_i=1$ if $X_i^m$ has been observed and $R_i=0$ otherwise).}
\label{exlin}
\begin{tabular}{ccc}
\hline
id & $R$ & $X^m$ \\
\hline
$i$ & 1 & $X_i^m$ \\
$i+1$ & 0 & $E(X^m|W_{i+1};\hat{\gamma}_n)$ \\
\hline \\
\end{tabular}
\end{table}

For example, consider the case of linear regression $Y=\theta'(1,Z_i')+\epsilon$ with incompletely observed dependent variable $Y$. In this case, estimation with complete data can be performed using the ordinary least squares method and the corresponding complete-data estimating equation is
\[
\Psi_n(\theta)=\frac{1}{n}\sum_{i=1}^n\left[Y_i-\theta'(1,Z_i')'\right](1,Z_i')'=0,
\]
where $Z_i$ is the vector of the independent variables for the $i$th observation. Therefore, the first stage in the estimation of $\theta_0$ with missing values in $Y$ proceeds by fitting a linear regression model for $E(Y|W)=\gamma_0'(1,W')'$ using the complete cases to obtain an estimate of $\gamma_0$ via least squares. The second stage of this analysis requires solving the estimating equation
\begin{eqnarray*}
\tilde{\Psi}_n(\hat{\theta}_n;\hat{\gamma}_n)&=&\frac{1}{n}\sum_{i=1}^n \left[R_iY_i+(1-R_i)\hat{\gamma}_n'(1,W_i')'-\hat{\theta}_n'(1,Z_i')'\right](1,Z_i')'=0.
\end{eqnarray*}
This equation leads to the usual closed-form least squares estimator for $\hat{\theta}_n$ where the missing $Y_i$ have been replaced by their estimated conditional expectation $\hat{\gamma}_n'(1,W_i')'$, given the fully observed variables $W$. In practice, this can be easily performed using standard software by replacing the missing $Y_i$s with the corresponding $\hat{\gamma}_n'(1,W_i')'$ and fitting a linear regression model in the resulting pseudo-complete dataset via least squares. Note that in this example we do not impose assumptions regarding the error distribution in the model the $Y_i=\gamma_0'(1,W_i')'+\epsilon_i$ since, if the mean model is correctly specified and $E(\epsilon)=0$, the least squares estimator $\hat{\gamma}_n$ is consistent for $\gamma_0$ and we just \textit{impute} the value $\hat{\gamma}_n'(1,W_i')'$ for the missing $Y_i$s. On the contrary, both Type A and Type B multiple imputation would impose strong distributional assumptions, such as normality, in order to simulate the missing $Y_i$s.

A related example is the case of a missing continuous covariate in linear regression. In this case one can assume a linear model for the incomplete covariate $Z_{ij}=\beta_0'(1,W_i')'+u_i$, where $E(u)=0$ and $\textrm{Var}(u|W_i)=\eta_0^2$. The parameter $\gamma_0$ is now $(\beta_0,\eta_0^2)$, where $\beta_0$ can be estimated via least squares and, then, $\eta_0^2$ as the sample average of $\hat{u}_i^2=(Z_{ij}-\hat{\beta}_n'(1,W_i')')^2$ using the compete cases. Since the estimating equation depends (linearly) on both $Z_{ij}$ and $Z_{ij}^2$, one can calculate the pseudo expected estimating equation by imputing the missing $Z_{ij}$ with
\[
E(Z_{ij}|W_i;\hat{\gamma}_n)=\hat{\beta}_n'(1,W_i')',
\]
and the missing $Z_{ij}^2$ with
\[
E(Z_{ij}^2|W_i;\hat{\gamma}_n)=[\hat{\beta}_n'(1,W_i')']^2+\hat{\eta}_n^2.
\]
Once again we were able to avoid distributional assumptions on the continuous incomplete variables. On the contrary, as mentioned above, both type A and type B multiple imputation would require distributional assumptions in order to simulate the missing $Z_{ij}s$, such as normality.

Another important example is the logistic regression model with incompletely observed dependent variable $Y$ and no auxiliary variables. In this case, the score function contribution for the $i$th observation is
\[
\psi_{\theta}(X_i)=\left[Y_i-\frac{\exp(\theta'(1,Z_i')')}{1+\exp(\theta'(1,Z_i')')}\right](1,Z_i')'
\]
where $Z_i$ is the vector of the independent variables for the $i$th observation. In this example, the incomplete variable is linear in $\psi_{\theta}(X_i)$. Therefore, estimation of $\theta_0$ proceeds by obtaining a preliminary estimate $\hat{\gamma}_n$ of $\gamma_0$ based on the complete cases and then by solving
\begin{eqnarray*}
\tilde{\Psi}_n(\hat{\theta}_n;\hat{\gamma}_n)&=&\frac{1}{n}\sum_{i=1}^n \left[R_iY_i+(1-R_i)\frac{\exp(\hat{\gamma}_n'(1,Z_i')')}{1+\exp(\hat{\gamma}_n'(1,Z_i')')}-\frac{\exp(\hat{\theta}_n'(1,Z_i')')}{1+\exp(\hat{\theta}_n'(1,Z_i')')}\right](1,Z_i')'=0,
\end{eqnarray*}
because under the logistic model
\[
E(Y|Z_i;\hat{\gamma}_n)=\frac{\exp(\hat{\gamma}_n'(1,Z_i')')}{1+\exp(\hat{\gamma}_n'(1,Z_i')')}.
\]

\subsection{Discrete Incomplete Variable with Finitely Many Values}
If $\psi_{\theta}(X^o,X^m)$ is not linear in the incomplete variable $X^m$, and $X^m$ is discrete taking its values in the finite set $\{x_1,\ldots,x_K\}$, then
\[
E[\psi_{\theta}(X^o,X^m)|W_i;\gamma_0]=\sum_{k=1}^K\psi_{\theta}(X_i^o,x_k)P(X^m=x_k|W_i;\gamma_0),
\]
for the missing cases. Estimation in this case can be easily performed in this case using standard software that allows for case weights. The first step is to create $K$ pseudo-records for each missing case in the dataset, one for each possible value of $X^m$. Next, based on an estimate $\hat{\gamma}_n$ from the complete cases, one needs to create a variable containing case weights equal to $P(X^m=x_k|W_i;\hat{\gamma}_n)$, $k=1,\ldots,K$, for the missing cases, and equal to 1 for the complete cases. An example of such a data setup is presented in Table~\ref{dataex}. Now, the analysis can be performed using standard software based on the augmented dataset, by weighting each observation using the aforementioned probability weights in the analysis.

\begin{table}
\centering
\caption{Example of the dataset up for two hypothetical individuals $i$ and $i+1$ in the case of a discrete incomplete variable $X^m$ with a finite set of values $\{x_1,\ldots,x_K\}$. ($R$: missingness indicator with $R_i=1$ if $X_i^m$ has been observed and $R_i=0$ otherwise; $\hat{p}_{i+1,k}^m=P(X^m=x_k|W_{i+1};\hat{\gamma}_n)$ for $k=1,\ldots,K$).}
\label{dataex}
\begin{tabular}{cccc}
\hline
id & $R$ & $X^m$ & weight \\
\hline
$i$ & 1 & $X_i^m$ & 1   \\
$i+1$ & 0 & $x_1$ & $\hat{p}_{i+1,1}^m$  \\
$\vdots$ & $\vdots$ & $\vdots$ & $\vdots$  \\
$i+1$ & 0 & $x_K$ & $\hat{p}_{i+1,K}^m$  \\
\hline \\
\end{tabular}
\end{table}

For example, consider the logistic regression model with an incompletely observed independent variable $Z^m$ with $P(Z^m\in\{1,2,3\})=1$. In this case, a multinomial logit model can be imposed for $Z^m$ such that
\[
P(Z^m=k|W_i;\gamma_0)=\frac{\exp(\gamma_{0,k}'(1,W_i')')}{1+\sum_{l=1}^2\exp(\gamma_{0,l}'(1,W_i')')}, \ \ \ \ k=1,2,
\]
and
\[
P(Z^m=3|W_i;\gamma_0)=\frac{1}{1+\sum_{l=1}^2\exp(\gamma_{0,l}'(1,W_i')')},
\]
where $\gamma_0=(\gamma_{0,1}',\gamma_{0,2}')'$. In this case, estimation of $\theta$ proceeds by obtaining an estimate $\hat{\gamma}_n$ based on the complete cases and then by using the estimating function
\begin{eqnarray*}
\tilde{\Psi}_n(\theta;\hat{\gamma}_n)&=&\frac{1}{n}\sum_{i=1}^n \Bigg\{R_i\left[Y_i-\frac{\exp(\theta'(1,Z_i')')}{1+\exp(\theta'(1,Z_i')')}\right](1,Z_i')'+ \\
&&(1-R_i)\sum_{k=1}^3\left[Y_i-\frac{\exp(\theta'\tilde{Z}_{ik})}{1+\exp(\theta'(1,\tilde{Z}_{ik}')')}\right](1,\tilde{Z}_{ik}')'\frac{\exp(\hat{\gamma}_{n,k}'(1,W_i')')}{1+\sum_{l=1}^2\exp(\hat{\gamma}_{n,l}'(1,W_i')')}\Bigg\},
\end{eqnarray*}
where $\tilde{Z}_{ik}$ is $Z_i$ with $Z_i^m$ being replaced by $k$, and $\hat{\gamma}_{n,3}\equiv 0$. The data augmentation trick described above can be used to solve $\tilde{\Psi}_n(\hat{\theta}_n;\hat{\gamma}_n)=0$ using standard software that allows for case weights in logistic regression.

\subsection{No Explicit Form for $E[\psi_{\theta}(X^o,X^m)|W=w;\gamma_0]$}
In many cases $\psi_{\theta}(X^o,X^m)$ is neither linear in the incomplete variable $X^m$ nor $X^m$ is discrete with finitely many values, and there is no explicit form for $E[\psi_{\theta}(X^o,X^m)|W=w;\gamma_0]$. In such cases, it is convenient to use Monte Carlo methods to numerically approximate $E[\psi_{\theta}(X^o,X^m)| W_i;\hat{\gamma}_n]$. Using this approach, each missing value is being simulated $S$ times from the assumed distribution $F_{X^m|W}(x|W=w;\hat{\gamma}_n)$ and then $E[\psi_{\theta}(X^o,X^m)|W_i;\hat{\gamma}_n]$ is approximated by
\[
\frac{1}{S}\sum_{j=1}^S\psi_{\theta}(X_i^o,\tilde{X}_i^m(U_{ij},\hat{\gamma}_n)),
\]
for the missing cases, where $\tilde{X}_i^m(U_{ij},\hat{\gamma}_n)=F_{X^m|W}^{-1}(U_{ij}|W_i;\hat{\gamma}_n)$, provided that the latter inverse exists, and $U_{ij}\sim U(0,1)$. Therefore, the approximate empirical estimating function becomes:
\[
\tilde{\Psi}_{n,S}(\theta;\hat{\gamma}_n)=\frac{1}{n}\sum_{i=1}^n \left[R_i\psi_{\theta}(X_i^o,X_i^m)+(1-R_i)\frac{1}{S}\sum_{j=1}^S\psi_{\theta}(X_i^o,\tilde{X}_i^m(U_{ij},\hat{\gamma}_n))\right].
\]
Obviously, the choice of $S$ will affect the precision of the resulting estimate as well as the computing time. The equation $\tilde{\Psi}_{n,S}(\hat{\theta}_{n,S};\hat{\gamma}_n)=0$ can be solved using a data augmentation trick similar to that for the case of a discrete incomplete variable with finitely many values. More precisely, for each case with missing $X^m$ one needs to create $S$ pseudo records which include the $S$ simulation realizations from $F_{X^m|W}(x|W_i;\hat{\gamma}_n)$. Then, a weight that is equal to $1/S$ for the missing cases and to 1 for the complete cases needs to be generated. An example of such a data setup is presented in Table~\ref{dataex2}. Finally, estimation of $\theta_0$ is performed in standard software that allows for case weights.

\begin{table}
\centering
\caption{Example of the dataset up for two hypothetical individuals $i$ and $i+1$ in the case where there is no explicit form for $E[\psi_{\theta}(X^o,X^m)|W=w;\gamma_0]$. ($R$: missingness indicator with $R_i=1$ if $X_i^m$ has been observed and $R_i=0$ otherwise; $\tilde{X}_{i+1}^m(U_{i+1,j},\hat{\gamma}_n)$: simulation realization from $F_{X^m|W}(x;W_{i+1},\hat{\gamma}_n)$ for $j=1,\ldots,S$).}
\label{dataex2}
\begin{tabular}{cccc}
\hline
id & $R^m$ & $X^m$ & weight \\
\hline
$i$ & 1 & $X_i^m$ & 1   \\
$i+1$ & 0 & $\tilde{X}_{i+1}^m(U_{i+1,1},\hat{\gamma}_n)$ & $1/S$  \\
$\vdots$ & $\vdots$ & $\vdots$ & $\vdots$  \\
$i+1$ & 0 & $\tilde{X}_{i+1}^m(U_{i+1,S},\hat{\gamma}_n)$ & $1/S$  \\
\hline \\
\end{tabular}
\end{table}

It is of note that Monte Carlo PEEE approach is equivalent to the type B multiple imputation approach \citep{Wang98,Robins00}. This is because the estimating function for the latter approach is equal to $S\tilde{\Psi}_{n,S}(\theta;\hat{\gamma}_n)$ and, therefore, both approaches yield the same estimate. However, the key difference here is that we provide a way to estimate $\hat{\theta}_{n,S}$ using a dataset of size $n+m(S-1)$ instead of a dataset of size $nS$ for the type B multiple imputation approach. This is due to the fact that only the records with a missing value are expanded $S$ times. Our computation approach can lead to a substantially faster implementation.

\section*{Appendix C: Simulation study 2: More pronounced model misspecification}
\renewcommand{\thesubsection}{C.\arabic{subsection}}
\setcounter{subsection}{0}
We conducted an additional series of simulation experiments to evaluate the performance of the PEEE approach and the proposed closed-form variance estimator for various degrees of misspecification of the model for the incomplete variable. Additionally, the performance of using a more flexible parametric model for the incomplete variable in such situations was also assessed. We considered two covariates of interest; $Z_1$ was a continuous covariate simulated from the standard normal distribution while $Z_2$ was a binary variable with $P(Z_2=1)=0.4$. Moreover, an auxiliary variable $A$ was simulated from the uniform distribution $U[0,5]$. The dependent variable, denoted as $Y$, was a continuous variable of the form
\[
Y=\beta_0^* + \beta_1Z_1 +\beta_2Z_2 + h(A) + \epsilon,
\]
where $(\beta_0^*,\beta_1,\beta_2)=(1,1,1)$ and $\epsilon$ was simulated from the $t$-distribution with 3 degrees of freedom. This distribution was chosen to illustrate the fact that the PEEE approach does not impose distributional assumptions on the incomplete variables. The different choices for the nonlinear function $h(\cdot)$ led to different misspecification scenarios for the model for the incomplete variable $Y$. The functions $h(\cdot)$ that we used in this series of simulations are depicted in Figure~1.

\begin{figure}[htb]
\begin{center}
\includegraphics{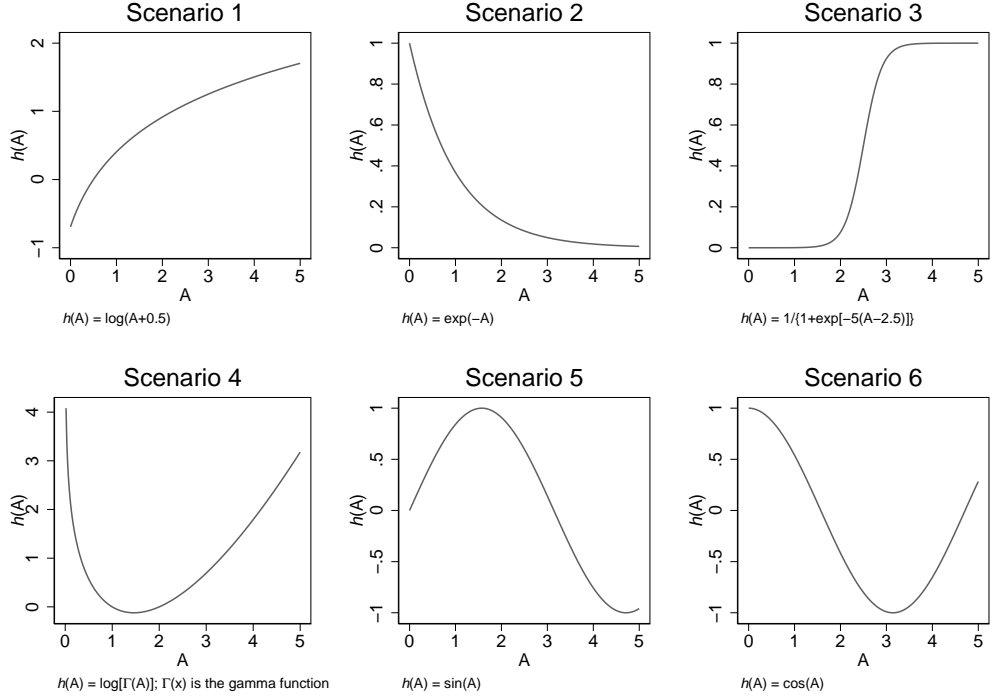}
\caption{True relationship between the response $Y$ and the auxiliary variable $A$ for each simulation scenario.}
\end{center}
\end{figure}

In these simulation studies, we considered missingness in the dependent variable $Y$. The missingness indicator was simulated based on the logit model
\[
\textrm{logit}(P(R=1|Z_1,Z_2,A))=-4.5 + Z_1+Z_2+2A,
\]
This simulation setup led on average to 41.3\% missingness. The auxiliary variable $A$ needs to be taken into account in order to satisfy the MAR assumption. This is because $A$ is associated with both the probability of missingess and the incompletely observed variable $Y$. Since $A$ is an auxiliary variable, the main model of interest is
\[
E(Y|Z_1,Z_2)=\beta_0+\beta_1Z_1+\beta_2Z_2,
\]
where $\beta_0=\beta_0^*+E[h(A)]$. In this simulation we evaluated the na\"ive compete case analysis, type B multiple imputation based on 10 imputations, and the PEEE approach along with the proposed closed-form variance estimator. The auxiliary variable was ignored in the complete case analysis. Two versions of the type B multiple imputation and the PEEE approach were considered. The first one used a misspecified simple linear \textit{imputation} model of the form
\[
E(Y|Z_1,Z_2,A)=\gamma_0^*+\gamma_1^*Z_1+\gamma_2^*Z_2+\gamma_3^*A,
\]
for the incomplete variable $Y$. The second version considered a more flexible \textit{imputation} model using regression B-splines of the form
\[
E(Y|Z_1,Z_2,A)=\gamma_0^*+\gamma_1^*Z_1+\gamma_2^*Z_2+\sum_{s=1}^{N+m}\gamma_{3,s}^*B_{s,m}(A),
\]
where $N$ and $m$ are the number of internal knots and the order of the B-spline, respectively. We set $N=3$ and $m=4$ which resulted in a cubic B-spline. Since the number of parameters in this model is fixed and does not grow with sample size, the proposed inference procedures for the PEEE approach are valid. It has to be noted that for the type B multiple imputation method, the missing values in $Y$ were simulated assuming a normal distribution instead of a $t$-distribution. Standard error estimation was based on the nonparametric bootstrap with 100 replications for the type B multiple imputation approach, and on the proposed closed-form estimator for the PEEE approach. Results from the simulation studies are listed in Tables~\ref{sim1}--\ref{sim6} .

\begin{table}
\caption{Simulation results under misspecification scenario 1 with $h(A)=\log(A+0.5)$, $n=5000$, and 41\% missingness.}
\label{sim1}
\centering
\begin{tabular}{ccccccc}
\hline
Method & $\beta$ & Bias (\%) & MCSD & ASE & CP & RE \\
\hline
CC&$\beta_0$&20.374&0.045&0.044&0.000&0.683\\
&$\beta_1$&-6.968&0.033&0.033&0.428&0.956\\
&$\beta_2$&-7.480&0.069&0.065&0.785&0.988\\[2ex]
MIB&$\beta_0$&4.714&0.055&0.053&0.583&1.015\\
&$\beta_1$&-0.373&0.034&0.034&0.945&1.008\\
&$\beta_2$&-0.905&0.070&0.067&0.943&1.033\\[2ex]
PEEE&$\beta_0$&4.713&0.054&0.053&0.579&1.000\\
&$\beta_1$&-0.366&0.033&0.033&0.946&1.000\\
&$\beta_2$&-0.924&0.069&0.066&0.942&1.000\\[2ex]
MIB(flex)&$\beta_0$&-0.046&0.067&0.064&0.950&1.510\\
&$\beta_1$&0.130&0.035&0.034&0.946&1.065\\
&$\beta_2$&-0.348&0.070&0.068&0.951&1.043\\[2ex]
PEEE(flex)&$\beta_0$&-0.042&0.067&0.063&0.945&1.503\\
&$\beta_1$&0.154&0.034&0.034&0.953&1.034\\
&$\beta_2$&-0.378&0.070&0.067&0.945&1.033\\
\hline
\end{tabular}
\vspace{1ex} \\
{\raggedright CC: Complete case analysis; MIB: Type B multiple imputation based on 10 imputations; 
MIB(flex): Type B multiple imputation based on 10 imputations and a flexible imputation model based on B-splines; PEEE: Proposed approach with a flexible model for the incomplete variable based on B-splines; MCSD: Monte Carlo standard deviation of the estimates; ASE: Average of the standard error estimates; CP: Empirical coverage probability of the 95\% confidence intervals; RE: Relative efficiency (larger values indicate larger variance compared to the variance of the PEEE approach). \par}
\end{table}

\begin{table}
\caption{Simulation results under misspecification scenario 2 with $h(A)=\exp(-A)$, $n=5000$, and 41\% missingness.}
\label{sim2}
\centering
\begin{tabular}{ccccccc}
\hline
Method & $\beta$ & Bias (\%) & MCSD & ASE & CP & RE \\
\hline
CC&$\beta_0$&-11.868&0.045&0.043&0.097&0.680\\
&$\beta_1$&2.116&0.032&0.032&0.900&0.955\\
&$\beta_2$&1.613&0.068&0.064&0.936&0.988\\[2ex]
MIB&$\beta_0$&-5.670&0.054&0.052&0.728&1.015\\
&$\beta_1$&0.499&0.033&0.033&0.943&1.013\\
&$\beta_2$&0.034&0.069&0.065&0.943&1.036\\[2ex]
PEEE&$\beta_0$&-5.672&0.054&0.052&0.725&1.000\\
&$\beta_1$&0.506&0.033&0.033&0.951&1.000\\
&$\beta_2$&0.015&0.068&0.065&0.946&1.000\\[2ex]
MIB(flex)&$\beta_0$&-0.069&0.066&0.064&0.946&1.479\\
&$\beta_1$&0.106&0.034&0.033&0.945&1.049\\
&$\beta_2$&-0.335&0.069&0.066&0.945&1.028\\[2ex]
PEEE(flex)&$\beta_0$&-0.061&0.066&0.062&0.945&1.475\\
&$\beta_1$&0.130&0.033&0.033&0.952&1.016\\
&$\beta_2$&-0.365&0.069&0.065&0.945&1.014\\
\hline
\end{tabular}
\vspace{1ex} \\
{\raggedright CC: Complete case analysis; MIB: Type B multiple imputation based on 10 imputations; 
MIB(flex): Type B multiple imputation based on 10 imputations and a flexible imputation model based on B-splines; PEEE: Proposed approach with a flexible model for the incomplete variable based on B-splines; MCSD: Monte Carlo standard deviation of the estimates; ASE: Average of the standard error estimates; CP: Empirical coverage probability of the 95\% confidence intervals; RE: Relative efficiency (larger values indicate larger variance compared to the variance of the PEEE approach). \par}
\end{table}

\begin{table}
\caption{Simulation results under misspecification scenario 3 with $h(A)=1/\{1+\exp[-5(A-2.5)]\}$, $n=5000$, and 41\% missingness.}
\label{sim3}
\centering
\begin{tabular}{ccccccc}
\hline
Method & $\beta$ & Bias (\%) & MCSD & ASE & CP & RE \\
\hline
CC&$\beta_0$&20.658&0.045&0.044&0.000&0.684\\
&$\beta_1$&-7.053&0.033&0.033&0.427&0.952\\
&$\beta_2$&-7.604&0.069&0.066&0.774&0.988\\[2ex]
MIB&$\beta_0$&1.515&0.055&0.053&0.921&1.015\\
&$\beta_1$&-0.834&0.033&0.034&0.948&1.008\\
&$\beta_2$&-1.403&0.070&0.067&0.939&1.034\\[2ex]
PEEE&$\beta_0$&1.513&0.054&0.053&0.921&1.000\\
&$\beta_1$&-0.828&0.033&0.033&0.945&1.000\\
&$\beta_2$&-1.422&0.069&0.066&0.945&1.000\\[2ex]
MIB(flex)&$\beta_0$&-0.100&0.067&0.064&0.950&1.497\\
&$\beta_1$&0.115&0.034&0.034&0.946&1.041\\
&$\beta_2$&-0.333&0.070&0.067&0.952&1.016\\[2ex]
PEEE(flex)&$\beta_0$&-0.094&0.067&0.063&0.946&1.492\\
&$\beta_1$&0.139&0.034&0.034&0.954&1.009\\
&$\beta_2$&-0.363&0.069&0.066&0.942&1.005\\
\hline
\end{tabular}
\vspace{1ex} \\
{\raggedright CC: Complete case analysis; MIB: Type B multiple imputation based on 10 imputations; 
MIB(flex): Type B multiple imputation based on 10 imputations and a flexible imputation model based on B-splines; PEEE: Proposed approach with a flexible model for the incomplete variable based on B-splines; MCSD: Monte Carlo standard deviation of the estimates; ASE: Average of the standard error estimates; CP: Empirical coverage probability of the 95\% confidence intervals; RE: Relative efficiency (larger values indicate larger variance compared to the variance of the PEEE approach). \par}
\end{table}

\begin{table}
\caption{Simulation results under misspecification scenario 4 with $h(A)=\log[\Gamma(A)]$, $n=5000$, and 41\% missingness.}
\label{sim4}
\centering
\begin{tabular}{ccccccc}
\hline
Method & $\beta$ & Bias (\%) & MCSD & ASE & CP & RE \\
\hline
CC&$\beta_0$&21.336&0.050&0.049&0.000&0.681\\
&$\beta_1$&-15.516&0.036&0.037&0.022&0.944\\
&$\beta_2$&-15.703&0.076&0.074&0.427&0.975\\[2ex]
MIB&$\beta_0$&-20.991&0.061&0.059&0.001&1.014\\
&$\beta_1$&2.496&0.038&0.038&0.893&1.005\\
&$\beta_2$&2.208&0.078&0.076&0.941&1.025\\[2ex]
PEEE&$\beta_0$&-20.992&0.061&0.059&0.001&1.000\\
&$\beta_1$&2.502&0.038&0.038&0.897&1.000\\
&$\beta_2$&2.189&0.077&0.075&0.934&1.000\\[2ex]
MIB(flex)&$\beta_0$&-0.155&0.070&0.067&0.941&1.308\\
&$\beta_1$&0.072&0.036&0.036&0.950&0.931\\
&$\beta_2$&-0.268&0.073&0.072&0.959&0.906\\[2ex]
PEEE(flex)&$\beta_0$&-0.150&0.070&0.066&0.938&1.304\\
&$\beta_1$&0.096&0.036&0.036&0.956&0.904\\
&$\beta_2$&-0.298&0.073&0.071&0.956&0.901\\
\hline
\end{tabular}
\vspace{1ex} \\
{\raggedright CC: Complete case analysis; MIB: Type B multiple imputation based on 10 imputations; 
MIB(flex): Type B multiple imputation based on 10 imputations and a flexible imputation model based on B-splines; PEEE: Proposed approach with a flexible model for the incomplete variable based on B-splines; MCSD: Monte Carlo standard deviation of the estimates; ASE: Average of the standard error estimates; CP: Empirical coverage probability of the 95\% confidence intervals; RE: Relative efficiency (larger values indicate larger variance compared to the variance of the PEEE approach). \par}
\end{table}

\begin{table}
\caption{Simulation results under misspecification scenario 5 with $h(A)=\sin(A)$, $n=5000$, and 41\% missingness.}
\label{sim5}
\centering
\begin{tabular}{ccccccc}
\hline
Method & $\beta$ & Bias (\%) & MCSD & ASE & CP & RE \\
\hline
CC&$\beta_0$&-35.411&0.048&0.046&0.000&0.700\\
&$\beta_1$&12.754&0.035&0.035&0.042&0.946\\
&$\beta_2$&12.081&0.073&0.069&0.570&0.989\\[2ex]
MIB&$\beta_0$&17.620&0.057&0.055&0.059&1.012\\
&$\beta_1$&-0.390&0.036&0.036&0.941&1.012\\
&$\beta_2$&-0.944&0.075&0.071&0.937&1.033\\[2ex]
PEEE&$\beta_0$&17.618&0.057&0.055&0.054&1.000\\
&$\beta_1$&-0.383&0.036&0.035&0.950&1.000\\
&$\beta_2$&-0.963&0.074&0.070&0.944&1.000\\[2ex]
MIB(flex)&$\beta_0$&-0.057&0.067&0.065&0.945&1.359\\
&$\beta_1$&0.136&0.035&0.035&0.940&0.969\\
&$\beta_2$&-0.378&0.072&0.069&0.947&0.961\\[2ex]
PEEE(flex)&$\beta_0$&-0.049&0.067&0.063&0.940&1.355\\
&$\beta_1$&0.160&0.035&0.034&0.949&0.942\\
&$\beta_2$&-0.408&0.072&0.068&0.946&0.945\\
\hline
\end{tabular}
\vspace{1ex} \\
{\raggedright CC: Complete case analysis; MIB: Type B multiple imputation based on 10 imputations; 
MIB(flex): Type B multiple imputation based on 10 imputations and a flexible imputation model based on B-splines; PEEE: Proposed approach with a flexible model for the incomplete variable based on B-splines; MCSD: Monte Carlo standard deviation of the estimates; ASE: Average of the standard error estimates; CP: Empirical coverage probability of the 95\% confidence intervals; RE: Relative efficiency (larger values indicate larger variance compared to the variance of the PEEE approach). \par}
\end{table}

\begin{table}
\caption{Simulation results under misspecification scenario 6 with $h(A)=\cos(A)$, $n=5000$, and 41\% missingness.}
\label{sim6}
\centering
\begin{tabular}{ccccccc}
\hline
Method & $\beta$ & Bias (\%) & MCSD & ASE & CP & RE \\
\hline
CC&$\beta_0$&-41.566&0.046&0.044&0.001&0.653\\
&$\beta_1$&3.516&0.033&0.033&0.826&0.962\\
&$\beta_2$&3.255&0.069&0.066&0.917&0.987\\[2ex]
MIB&$\beta_0$&-41.610&0.057&0.054&0.002&1.016\\
&$\beta_1$&3.515&0.034&0.034&0.834&1.013\\
&$\beta_2$&3.281&0.071&0.067&0.909&1.035\\[2ex]
PEEE&$\beta_0$&-41.614&0.057&0.054&0.002&1.000\\
&$\beta_1$&3.521&0.034&0.034&0.824&1.000\\
&$\beta_2$&3.261&0.070&0.067&0.912&1.000\\[2ex]
MIB(flex)&$\beta_0$&-0.136&0.066&0.065&0.944&1.377\\
&$\beta_1$&0.073&0.035&0.035&0.954&1.050\\
&$\beta_2$&-0.316&0.071&0.069&0.948&1.037\\[2ex]
PEEE(flex)&$\beta_0$&-0.125&0.066&0.063&0.942&1.375\\
&$\beta_1$&0.097&0.034&0.034&0.955&1.018\\
&$\beta_2$&-0.347&0.071&0.068&0.946&1.024\\
\hline
\end{tabular}
\vspace{1ex} \\
{\raggedright CC: Complete case analysis; MIB: Type B multiple imputation based on 10 imputations; 
MIB(flex): Type B multiple imputation based on 10 imputations and a flexible imputation model based on B-splines; PEEE: Proposed approach with a flexible model for the incomplete variable based on B-splines; MCSD: Monte Carlo standard deviation of the estimates; ASE: Average of the standard error estimates; CP: Empirical coverage probability of the 95\% confidence intervals; RE: Relative efficiency (larger values indicate larger variance compared to the variance of the PEEE approach). \par}
\end{table}

The simulation results revealed that the na\"ive complete case analysis can provide seriously biased estimates when there are auxiliary variables not accounted in the analysis. Both type B multiple imputation and the PEEE approach based on a misspecified linear model for the incomplete variable $Y$ also provided biased estimates for $\beta_0$. The bias was larger when the degree of model misspecification was more pronounced. Moreover, these methods provided estimates for $\beta_1$ and $\beta_2$ with a small bias (bias range: 2\% to 4\%) in scenarios 4 and 6. However, due to the very small standard error (owing to the large sample size considered), this small bias led to empirical coverage probabilities for $\beta_1$ and $\beta_2$ with a lower coverage rate in these scenarios. It has to be noted that the proposed standard error estimates were close to the Monte Carlo standard deviation of the estimates in all cases. The same behavior was observed for the bootstrap standard errors for type B multiple imputation based on 100 replications. The PEEE approach was slightly more efficient compared to type B multiple imputation based on 10 imputations. Using a more flexible model  for the incomplete variables based on B-splines led to virtually unbiased estimates and coverage probabilities close to the nominal level for both PEEE and type B multiple imputation in all cases. However, adopting a more flexible model led to an inflation of the standard errors. This phenomenon was slightly less pronounced for the PEEE approach. Interestingly, simulating the missing values from a normal distribution instead of the true $t$-distribution with 3 degrees of freedom did not affect the performance of type B multiple imputation. To sum up, the simulation results indicate that, under misspecification of the incomplete variable model, the proposed closed-form variance estimator performs well and that the PEEE estimator behaves similarly to multiple imputation based on a parametric imputation model. A useful remedy for a possible model misspecification is to adopt a more flexible model for the incomplete variable.

\section*{Appendix D: R code for the analysis of the motivating data\label{app2}}
\renewcommand{\thesubsection}{D.\arabic{subsection}}
\setcounter{subsection}{0}

The analysis of the motivating TBI dataset can be analyzed using a simple data management step along with a couple of auxiliary functions. The first step of the analysis is to create the response and missingness indicators as
\begin{verbatim}
data$r <- !is.na(data$race)
\end{verbatim}
where \texttt{data} is the dataset name and \texttt{race} the incomplete three-level race variable. Next, for each patient with missing race, three pseudo-records are created using the following code:
\begin{verbatim}
dat0 <- data[data$r==1,]
dat0$race_dup <- dat0$race
dat1 <- data[data$r==0,]
dat1$race_dup <- 1
dat2 <- data[data$r==0,]
dat2$race_dup <- 2
dat3 <- data[data$r==0,]
dat3$race_dup <- 3
dat <- rbind(dat0, dat1, dat2, dat3)
\end{verbatim}
The variable \texttt{race\_dup} includes either the observed race or the possible race values for the missing cases. Next, we need to define the function that fits a multinomial logit model for to the complete cases with \texttt{race} as the dependent variable and \texttt{male} (male status indicator), \texttt{age} (age in years), \texttt{er} (emergency room prior to TBI diagnosis), \texttt{rehab} (rehabilitation within three months from hospitalization) and \texttt{year} (auxiliary variable of calendar year of hospitalization) as independent variables. The function needs to return a weight equal to 1 for the observed cases (i.e. cases with \texttt{data\$r==1}) and equal to the predicted probability for the corresponding value of \texttt{race\_dup} for the missing cases (i.e. cases with \texttt{data\$r==0}).
First, we need to load the package \texttt{nnet} which provides the function \texttt{multinom} that fits multinomial logit models.
\begin{verbatim}
library(nnet)
\end{verbatim}
We also need to load the following packages

\begin{verbatim}
library(sandwich)
library(tidyverse)
library(MASS)
\end{verbatim}
Next, we need to define the function that provides the weights, the estimated coefficients $\hat{\gamma}_n$ and the estimated variance matrix for $\hat{\gamma}_n$ as follows:

\begin{verbatim}
wt_include = function(dat){
  mod= multinom(race ~ male + age + er + rehab + year, data=dat)
  gamma = summary(mod)$coefficients
  pred.prob = predict(mod, newdata = dat, type = "probs")
  wt.data = as.data.frame(pred.prob)
  names(wt.data) = c("wt1", "wt2", "wt3")
  dat = cbind(dat, wt.data)
  dat2 = within(dat,{
    wt = (race_dup==1)*wt1 + (race_dup==2)*wt2 + (race_dup==3)*wt3
    wt = ifelse(!is.na(race), 1, wt)
  })

  drops <- c("race","wt1","wt2","wt3")
  dat2 <- dat2[ , !(names(dat2) %in% drops)]

  list(data= dat2, gamma=gamma, V_gamma = vcov(mod))
}
\end{verbatim}
The next function is used to provide the PEEE estimates $\hat{\beta}_n$ using the dataset with the weights from the \texttt{wt\_include} function, along with quantities that are being used in the closed-form variance estimator.

\begin{verbatim}
peee_logit = function(dat){
  fit = glm(rehab ~ male + age + as.factor(race_dup) + er,
            weights = wt,
            family="binomial",
            data=dat)
  resid =  residuals(fit, type="response") # fitted residuals

  # output: 1) regression coeff; 2) fitted residuals;
  ## 3) observed variance covariance matrix; 4) individual score contributions
  out = list(beta= coefficients(fit),
             resids = resid,
             V_beta = vcov(fit),
             psi=estfun(fit)
  )
  out
}
\end{verbatim}
To calculate the predicted probabilities of each race we need the auxiliary functions:

\begin{verbatim}
expit1 = function(x1, x2) exp(x1)/(1 + exp(x1) + exp(x2))
expit2 = function(x1, x2) exp(x2)/(1 + exp(x1) + exp(x2))
\end{verbatim}
A key element in variance estimation is the calculation of
\[
\hat{G}_{\hat{\theta}_n,\hat{\gamma}_n}=\frac{1}{n}\sum_{i=1}^n\left[(1-R_i)\frac{\partial}{\partial\gamma}E[\psi_{\hat{\theta}_n}(X_i)|W_i;\gamma]\big|_{\gamma=\hat{\gamma}_n}\right].
\]
For this we will utilize the numerical differentiation function \texttt{fdjac}:
\begin{verbatim}
const= sqrt(.Machine$double.neg.eps)
fdjac <- function(PARAM,G,...,eps=const) {
  # Computes a finite difference approximation to the Jacobian of G
  # G is a (possibly vector valued) function
  # PARAM is the point where the Jacobian is approximated
  # ... = additional arguments to G
  # eps = step size in finite difference approximation
  # (scaled by max(abs(PARAM[j]),1))
  N <- length(PARAM)
  G0 <- G(PARAM,...)
  JAC <- matrix(0,length(G0),N)
  for (j in 1:N) {
    X1 <- PARAM
    X1[j] <- PARAM[j] + eps*max(abs(PARAM[j]),1)
    JAC[,j] <- (G(X1,...)-G0)/(X1[j]-PARAM[j]) #divide by actual difference
  }
  JAC
}
\end{verbatim}
The \texttt{fdjac} function will be utilized in the function below which provides $\hat{G}_{\hat{\theta}_n,\hat{\gamma}_n}$:
\begin{verbatim}
jac = function(dat, resid, gm) {
  analysis_score = function(gamma2, gamma3) {
    gamma2 = matrix(gamma2, ncol = 1)
    gamma3 = matrix(gamma3, ncol = 1)
    Z = model.matrix( ~ male + age + as.factor(race_dup) + er,
                      data = dat) # analysis model
    W = model.matrix( ~ male + age + er + rehab + year,
                       data = dat) # incomplete variable model

    wts = (
      as.numeric(dat$r == 1) +
        as.numeric(dat$r == 0) * as.numeric(dat$race_dup == 2) *
          expit1(x1 = W %*% gamma2, x2 = W %*% gamma3) +
        as.numeric(dat$r == 0) * as.numeric(dat$race_dup == 3) *
          expit2(x1 = W %*% gamma2, x2 = W %*% gamma3) +
        as.numeric(dat$r == 0) * as.numeric(dat$race_dup == 1) *
          (1 - expit1(x1 = W %*% gamma2, x2 = W %*% gamma3)-
             expit2(x1 = W %*% gamma2, x2 = W %*% gamma3))
    )
    comp2 = matrix((1 - dat$r) * as.vector(wts) * resid, ncol = 1)
    out = (1/sum(wts))*t(Z) %*% comp2
    out
  }

  # derivative with respect to gamma
  deri_wrt_gm = fdjac(PARAM = gm,
                      function(v.tmp)
            analysis_score(gamma2 = v.tmp[1:(length(v.tmp) / 2)],
            gamma3 = v.tmp[(length(v.tmp) /2 + 1):length(v.tmp)]))

  deri_wrt_gm
}
\end{verbatim}
The final auxiliary function needed is used to extract the individual estimating function contributions for the incomplete variable multinomial logit model. Note that for other generalized linear models, the function \texttt{estfun} of the R package \texttt{sandwich} can be used to automatically calculate the individual estimating function contributions, as we did in the \texttt{peee\_logit} function defined above. However, the \texttt{estfun} function currently does not work with multinomial logit models. For this we develop the \texttt{estfun\_multinom} function:
\begin{verbatim}
estfun_multinom = function(dat, gamma2, gamma3){
  gamma2 = matrix(gamma2, ncol=1)
  gamma3 = matrix(gamma3, ncol=1)
  W = model.matrix(~ male + age + er + rehab + year, data=dat)

  wt1 = as.vector( -1*
  (as.numeric(!(dat$race_dup==2))*expit1(x1=W%*%gamma2,x2= W%*%gamma3)) +
  (as.numeric(dat$race_dup==2)*(1 -expit1(x1=W%*%gamma2,x2= W%*%gamma3)))
  )

  wt2 = as.vector( -1*(as.numeric(!(dat$race_dup==3))*
        expit2(x1=W%*%gamma2,x2= W%*%gamma3)) +
        (as.numeric( dat$race_dup==3)*
        (1 -expit2(x1=W%*%gamma2,x2= W%*%gamma3)))
  )

  comp1 = W*(wt1*dat$r)
  comp2 = W*(wt2*dat$r)
  score_contri = cbind(comp1, comp2)
  score_contri
}
\end{verbatim}
Finally, the function for providing the estimated variance matrix of $\hat{\beta}_n$, which is $n^{-1}\hat{\Omega}_n$, is
\begin{verbatim}
Omega = function(dat, fit, V_gamma, score_miss_model, G_hat){
  V_beta = fit$V_beta

  U_mod_a = fit$psi
  U_mod_a = as.data.frame(U_mod_a)
  U_mod_a$id = dat$id
  U_mod_a = tbl_df(U_mod_a)
  U_mod_b  = U_mod_a%>%
    group_by(id) %>%
    summarise(V1= sum(`(Intercept)`), V2=sum(male),
              V3 = sum(age),
              V4 = sum(`as.factor(race_dup)2`),
              V5 = sum(`as.factor(race_dup)3`),
              V6 = sum(er))
  U_mod = as.data.frame(U_mod_b)
  U_mod = U_mod[,!names(U_mod) %in% c("id")]
  psi = as.matrix(U_mod)
  U_miss_a = score_miss_model
  U_miss_a  = as.data.frame(U_miss_a)
  names(U_miss_a) = paste("V",1:ncol(U_miss_a),sep="")

  U_miss_a$id = dat$id
  U_miss_a = tbl_df(U_miss_a)
  U_miss_b   = U_miss_a %>%
    group_by(id) %>%
    summarise(V1= sum(V1), V2=sum(V2),
              V3= sum(V3), V4=sum(V4),
              V5= sum(V5), V6=sum(V6),
              V7= sum(V7),  V8=sum(V8),
              V9= sum(V9),  V10=sum(V10),
              V11= sum(V11),  V12=sum(V12))

  U_miss = as.data.frame(U_miss_b)
  U_miss  =  U_miss[,!names(U_miss) %in% c("id")]
  U_miss = as.matrix(U_miss)

  #Influence function for the gamma estimator
  I_gamma = V_gamma*nrow(U_mod)
  omega = I_gamma%*%t(U_miss)

  Psi_dot_inv = V_beta*nrow(U_mod)
  #Proposed sandwich-type variance estimator
  Omega = (1/nrow(U_mod))*Psi_dot_inv%*%(t(U_mod) + G_hat%*%omega)%*%
    t(t(U_mod) + G_hat%*%omega)%*%Psi_dot_inv
  Var = Omega/nrow(U_mod)
  return(Var)
}
\end{verbatim}
Now, the analysis of the augmented dataset \texttt{dat} can be easily performed as follows:

\begin{verbatim}
# First generate the weights and get the PEEE estimates
dat.peee <- wt_include(dat=dat)
fit <- peee_logit(dat.peee[[1]])

# Next calculate G_hat and the estimating functions for the
# multinomial model for the incomplete race
G_hat = jac(dat = dat.peee[[1]],
            resid = fit$resids,
            gm = c(dat.peee$gamma[1,],
                   dat.peee$gamma[2,]))

score_miss_model = estfun_multinom(dat = dat.peee[[1]],
                                  gamma2 = dat.peee$gamma[1,],
                                  gamma3 = dat.peee$gamma[2,])

#Finally calculate the variance matrix
Sigma <- Omega(dat = dat.peee$data, fit = fit,
                 score_miss_model = score_miss_model,
                 G_hat = G_hat)
\end{verbatim}

\end{appendices}

\end{document}